\documentclass[a4paper,fleqn,usenatbib]{mnras}

\usepackage{pdflscape}

\usepackage[T1]{fontenc}
\usepackage{ae,aecompl}

\usepackage{graphicx}	
\usepackage{amsmath}	
\usepackage{amssymb}	

\title[Major CRTS AGN flares]{Understanding extreme quasar optical variability with CRTS: I. Major AGN flares }

\author[M. J. Graham et al.]{
Matthew J. Graham,$^{1}$\thanks{E-mail:mjg@caltech.edu (MJG)} 
S.~G.~Djorgovski,$^1$ 
Andrew~J.~Drake,$^1$ 
Daniel Stern,$^2$ 
\newauthor
Ashish~A.~Mahabal,$^1$  
Eilat~Glikman,$^3$ 
Steve Larson$^4$
and Eric Christensen$^4$
\\
$^{1}$California Institute of Technology, 1200 E. California Blvd, Pasadena, CA 91125, USA \\
$^{2}$Jet Propulsion Laboratory, California Institute of Technology, 4800 Oak Grove Drive, Pasadena, CA 91109, USA \\
$^{3}$Department of Physics, Middlebury College, Middlebury, VT 05753, USA\\
$^{4}$University of Arizona, Department of Planetary Sciences, Lunar and Planetary Lab, Tucson, AZ 85721, USA
}

\date{Accepted 2017 June 08. Received 2017 June 06; in original form 2017 March 08}

\pubyear{2017}

\begin{document}
\label{firstpage}
\pagerange{\pageref{firstpage}--\pageref{lastpage}}
\maketitle

\begin{abstract}
There is a large degree of variety in the optical variability of quasars and it is unclear whether this is all attributable to a single (set of) physical mechanism(s). We present the results of a systematic search for major flares in AGN in the Catalina Real-time Transient Survey as part of a broader study into extreme quasar variability. Such flares are defined in a quantitative manner as being atop of the normal, stochastic variability of quasars. We have identified 51 events from over 900,000 known quasars and high probability quasar candidates, typically lasting 900 days and with a median peak amplitude of $\Delta m = 1.25$ mag. Characterizing the flare profile with a Weibull distribution, we find that nine of the sources are well described by a single-point single-lens model. This supports the proposal by Lawrence et al. (2016) that microlensing is a plausible physical mechanism for extreme variability. However, we attribute the majority of our events to explosive stellar-related activity in the accretion disk: superluminous supernovae, tidal disruption events, and mergers of stellar mass black holes. 
\end{abstract}




\section{Introduction}

Quasars are known to be a variable population, best described {\em statistically} via a damped random walk (DRW) model, (e.g., \cite{kelly09}, although see \cite{kozlowski16} for a reappraisal). Their observed optical variability is typically a few tenths of a magnitude in amplitude with a characteristic timescale of several months, but also showing larger variations over longer timescales \citep{macleod12,macleod16,kozlowski16}. The variability amplitude and timescale are anti-correlated with both luminosity and Eddington ratio, and correlated with black hole mass. The (extreme) tails of the variability distribution are less well characterized, however. While it is well established that blazars tend to exhibit large amplitude, short timescale variability, large amplitude variability on longer timescales is not as well studied, and seems to probe a different population than blazars. 

We have previously reported on a set of quasars showing periodic variability which is consistent with a population of supermassive black hole binaries with sub-parsec separation \citep{graham15a, graham15b, dorazio15} . Recently a number of objects - so-called changing look quasars \citep{ruan16, macleod16, lamassa15} - have been reported, showing slow but consistent photometric variability ($\Delta$m $> 1$ mag) over several years coupled with spectral variability. Their optical spectra show emerging or disappearing broad emission line components, typically H$\beta$. This is consistent with a change of type (Type 1 - Type 1.2/1.5 to Type 1.8/1.9 - Type 2 or vice versa) and may be associated with a large change of obscuration or accretion rate. We have also reported a more extreme example of photometric and spectral variability exhibited by a BAL QSO, most probably experiencing a change in photoionization (\cite{stern17}, see also \cite{rafiee16}).

Clearly a much larger sample of extreme variable quasars is needed to fully understand the different physical mechanisms that may be contributing to the observed phenomena and also to determine whether or how they relate to variability seen in the more general quasar population. Extreme quasars are, by definition, rare but the growing availability of large archives of astronomical time series, e.g., SDSS Stripe 82 \citep{sesar07}, LINEAR \citep{sesar11}, PTF \citep{rau09}, and Pan-STARRS \citep{kaiser04}, means that statistically valid samples of such objects can now be defined. 

We have begun a systematic study of extreme quasar variability based primarily on the Catalina Real-time Transient Survey (CRTS\footnote{http://crts.caltech.edu}; \cite{drake09,mahabal11,djorgovski12}). This is the largest open (publicly accessible) time domain survey currently operating, covering $\sim 33,000$ deg$^2$ in the range $-75^{\circ} < \mathrm{Dec} < 70^{\circ}$ (but avoiding regions within $\sim 10^{\circ} - 15^{\circ}$ of the Galactic plane) to a depth of $V \sim 19$ to 21.5. Time series exist\footnote{http://www.catalinadata.data} for approximately 500 million objects with an average of $\sim$300 observations over a 11-year baseline. 

In this work, we present a search for major flaring outbursts in AGN. Subsequent papers will deal with other phenomena exhibited by extreme variable quasars (such as discussed above). Although there have been several reports of significant optical/UV outbursts in quiescent galaxies, consistent with superluminous supernovae or candidate tidal disruption events (TDEs) \citep{gezari12, chornock14, komossa15}, those associated with active galaxies are much rarer as it can be difficult to distinguish a single significant event from more general variability.
They have the potential, however, to inform about the structure and mechanics of the accretion disk and nuclear region.

Descriptions of major AGN flaring outbursts in the literature to date tend to deal with individual events. \cite{meusinger10} reported a significant UV flare in a quasar, Sharov 21 at $z = 2.109$, seen through the disk of M31 (this had previously been misidentified as a nova event). The total outburst lasted $\sim$800 days with the source 3.3 mag brighter at maximum. The flare showed an asymmetric profile with a gradual increase followed by an abrupt rise and then a quasi-exponential decline $(t^{-5/3})$ and a total bolometric energy release of $\sim 2 \times 10^{54}$ ergs. It is consistent with a standard TDE scenario involving a $\sim 10 M_{\odot}$ star and a $5 \times 10^8 M_\odot$ black hole (although microlensing is also considered as an alternate explanation). The TDE explanation, however, neglects the AGN nature of the host and the influence of the massive accretion disk and general relativistic effects on the dynamics of the stellar tidal debris.

\cite{drake11} discovered an extremely luminous optical transient within 150 pc of the nucleus of a narrow-line Seyfert 1 galaxy, SDSS J102912+404220 at $z = 0.147$. The total outburst lasted $\sim$400 days with the source $\sim$1.2 mag brighter at maximum.  It also showed an asymmetric profile with a slow increase and then a longer slow decline that was inconsistent with the expected $t^{-5/3}$ or $t^{-5/12}$ decline expected for TDEs. The proposed interpretation is an extremely luminous Type IIn supernova within the range of the narrow-line region of an AGN.

\cite{lawrence16} (hereafter L16) reported a search for large amplitude ($\Delta m > 1.5$ mag) nuclear changes in faint extragalactic objects in the PS1 3$\pi$ survey compared against SDSS data over 11,663 deg$^2$. Of the 76 transients detected, 43 are classed as ``slow blue hypervariable'' AGN with smooth order of magnitude outbursts over several years, large colour changes between the SDSS and PS1 epochs, and weaker than average broad emission line strength in their spectra. A combination of changes in accretion state and large amplitude microlensing by stars in foreground galaxies seem to be the most likely explanations. \cite{bruce16} have also reported a more detailed analysis of four of the lensing candidates, with two well described by a simple single point-lens point-source model and the other two requiring a more complex lensing model. Although microlensing has the potential for mapping the inner structure of an AGN, these events only place minor constraints on the size of \ion{C}{iii}] and \ion{Mg}{ii} emission regions.

This paper is structured as follows: in section 2, we present the selection technique for identifying major flaring activity and in section 3, the data sets we have applied it to. We discuss our results in section 4 and their interpretation in section 5. We assume a standard WMAP 9-year cosmology ($\Omega_\Lambda = 0.728$, $\Omega_M = 0.272$, $H_0 = 70.4$ km s$^{-1}$ Mpc$^{-1}$; \cite{jarosik}) and our magnitudes are approximately  on the Vega system.

\section{Data sets}
There are few data sets with sufficient sky and/or temporal coverage and sampling to support an extensive search for quasars exhibiting significant flaring. Most large studies of long-term quasar variability, e.g.,
SDSS with POSS \citep{macleod12} or Pan-STARRS1 \citep{morganson14}, consist of relatively few epochs of data spread over a roughly decadal baseline,  which is sufficient to model ensemble behavior but not to identify specific patterns in individual objects beyond a change of magnitude. CRTS represents the best data currently available with which to systematically define sets of quasars with particular temporal characteristics.

\subsection{Catalina Real-time Transient Survey (CRTS)}
\label{crts}

CRTS leverages the Catalina Sky Survey data streams from three telescopes --  the 0.7 m Catalina Sky Survey Schmidt and 1.5 m Mount Lemmon Survey telescopes in Arizona and  the 0.5 m Siding Springs Survey Schmidt in Australia -- used in a search for Near-Earth Objects, operated by Lunar and Planetary Laboratory at University of Arizona. CRTS covers up to $\sim$2500 deg$^2$ per night, with 4 exposures per visit, separated by 10 min, over 21 nights per lunation. New cameras in Fall 2016 with larger fields-of-view will increase the nightly sky coverage. All data are automatically processed in real-time, and optical transients are immediately distributed using a variety of electronic mechanisms\footnote{http://www.skyalert.org}. The data are unfiltered but are broadly calibrated to Johnson $V$ from 2MASS data (see \citealt{drake13} for details).
The accuracy of the $V$-band photometry is highly dependent on source colour but comparison with Landolt standard stars has shown that the colour correction is small for blue objects. The effect on quasar variability should therefore be minimal. The full CRTS data set\footnote{http://nesssi.cacr.caltech.edu/DataRelease} contains time series for approximately 500 million sources. 

\begin{figure}
\centering
\caption{Corrective factor for photometric errors in CSS}
\label{errcorr}
\includegraphics[width=3.3in]{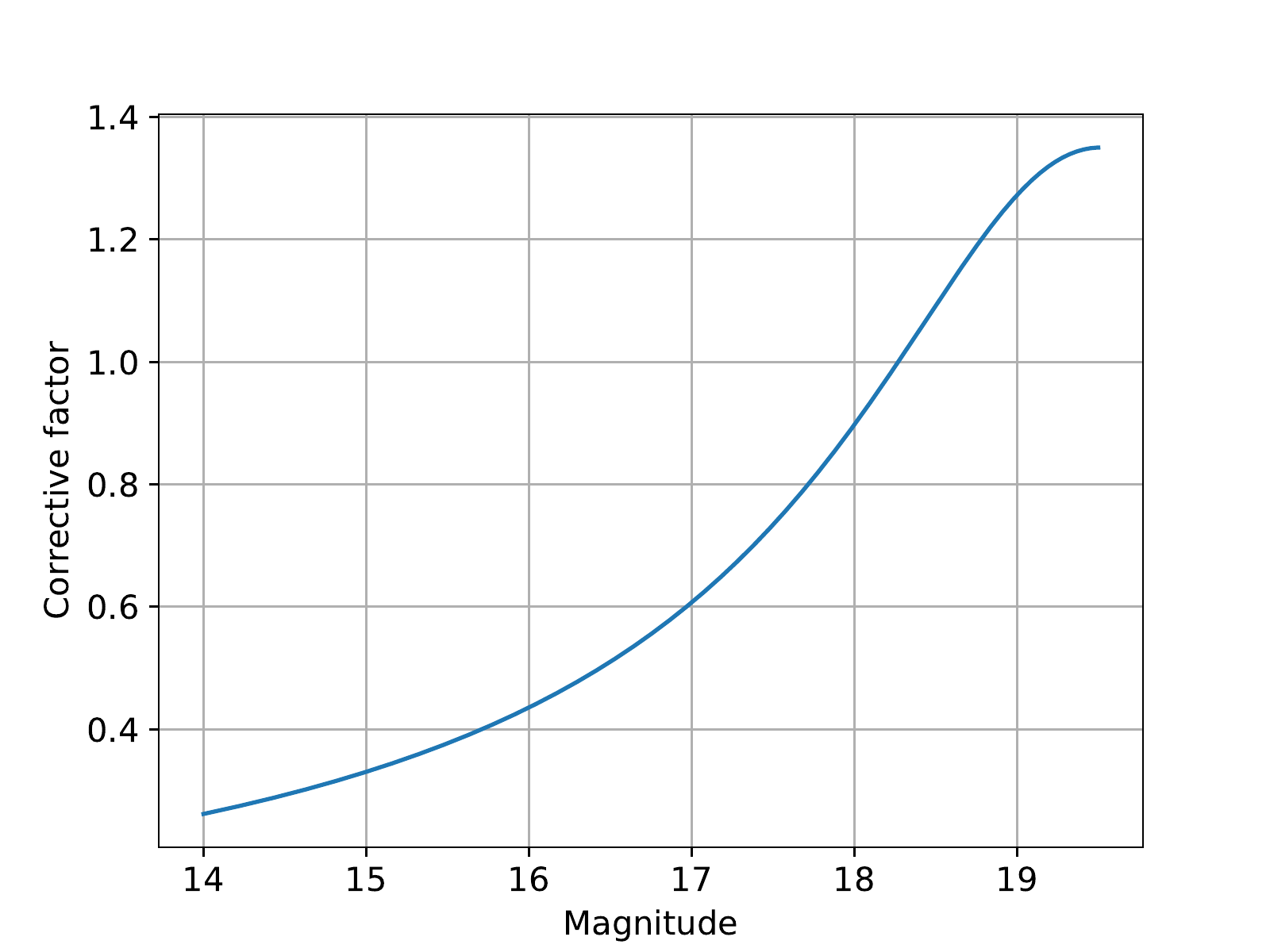}
\end{figure}

We note that the published error model for CRTS is incorrect. The photometric uncertainties were originally determined
\footnote{http://nesssi.cacr.caltech.edu/Data/FAQ2.html\#uncert} via an empirical relationship between source flux and the observed photometric scatter. 
This relation was derived from 100,000 isotropically selected sources that exhibited no significant sign of variability based on their Welch-Stetson variability index. However, errors at the brighter magnitudes are overestimated and those are fainter magnitudes ($> 18$) are underestimated \citep{palaversa13, drake14, vaughan16}. We have derived a multiplicative correction factor from CRTS observations of 350,000 sources in the Stripe 82 Standard Star catalogue \cite{ivezic07} that ensures that the mode of the reduced chi-squared variability in magnitude bins of width of $\Delta$ mag $ = 0.05$ is centred at unity (see Fig.~\ref{errcorr}).

\begin{figure}
\centering
\caption{Sky coverage (RA, Dec, Mollweide projection) for sources considered in this paper. The colours represent the subsamples: known QSOs (red), XDQSO (blue), WISE (green) and variables (grey).}
\label{coverage}
\includegraphics[width = 3.3in] {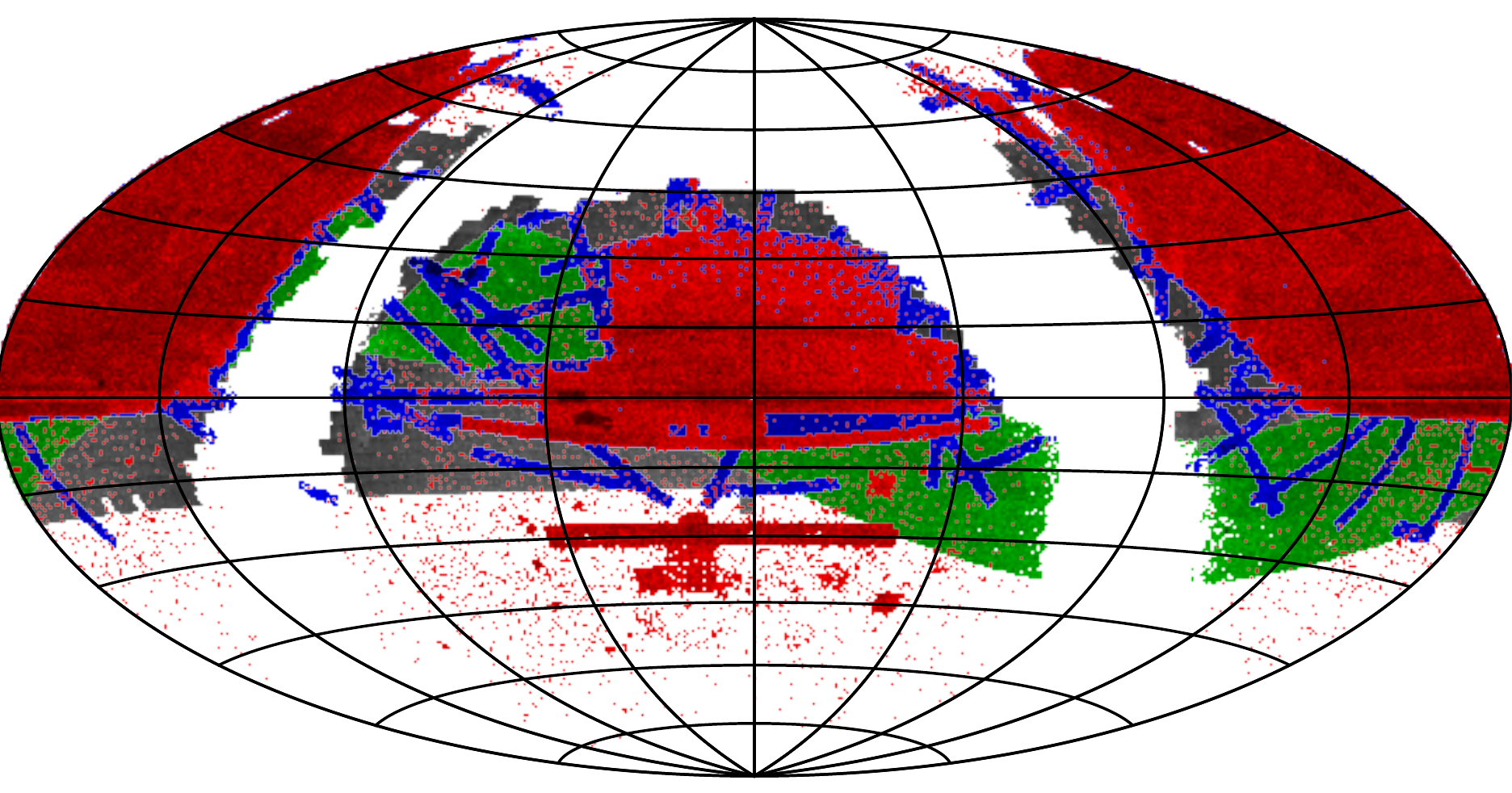}
\end{figure}

\begin{figure}
\centering
\caption{Relative $V$ magnitude distributions for sources considered in this paper. The colours represent known QSOs (red), XDQSO (blue), WISE (green) and variables (black).}
\label{totalmag}
\includegraphics[width = 2.8in] {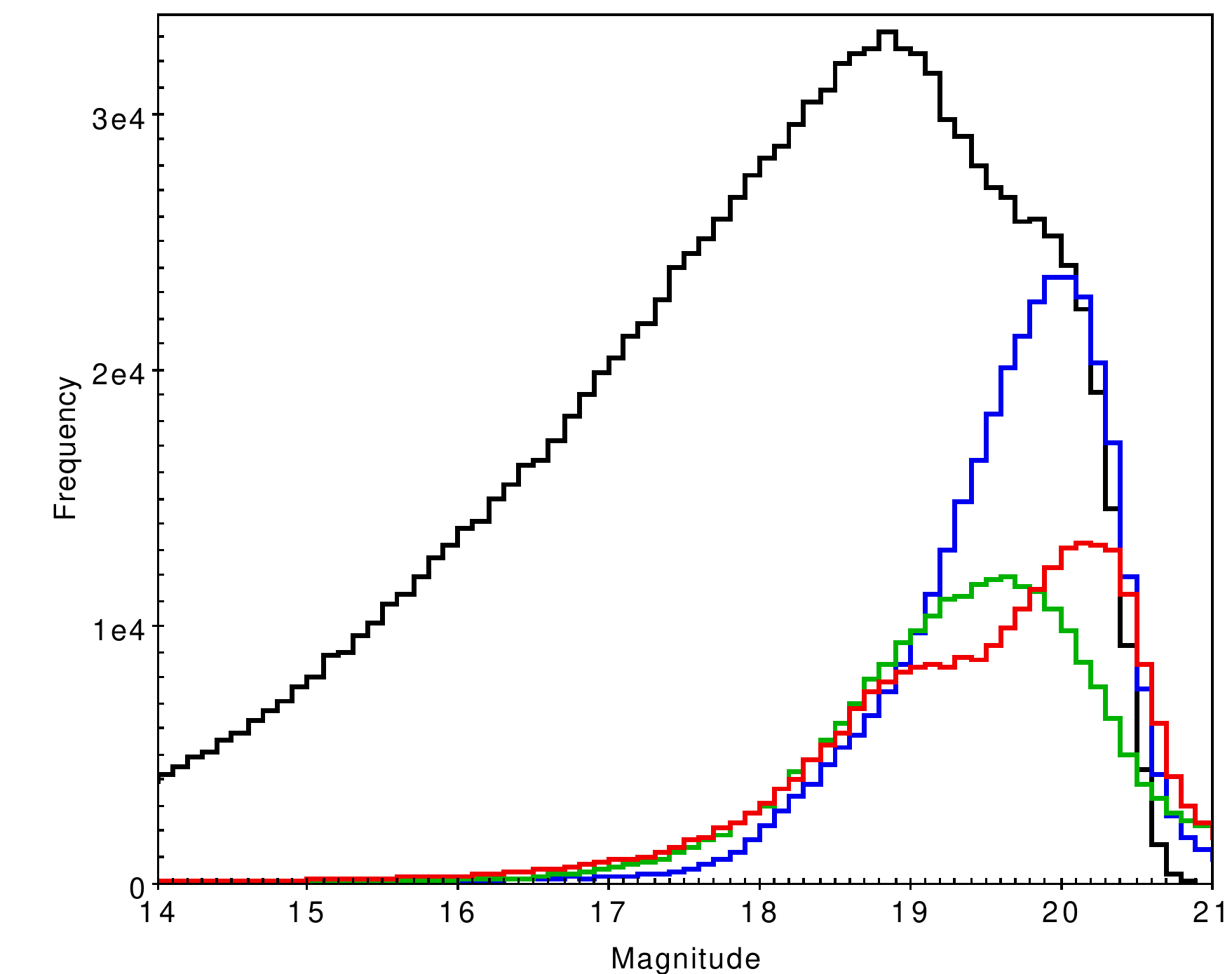}
\end{figure}

We have extracted a number of samples from CRTS in which to search for significant AGN flaring and these are summarized in Table~\ref{datasets}. Repeat instances of sources between different samples are ignored, i.e, they are included in the first sample but not subsequent ones. The total number of unique sources employed is 2,127,266 and their sky coverage and $V$ magnitude distribution are shown in Figs.~\ref{coverage} and \ref{totalmag} respectively. We also note that none of these sources have fewer than 10 observations in their light curve. We apply the same preprocessing steps described in \cite{graham15b} to all light curves.

\begin{table*}
  \centering
  \caption{List of CRTS samples searched for significant AGN flaring activity. The number of unique sources refers to objects only appearing in that data set -- repeat instances in other data sets are ignored; e.g., a source identified both as a known spectroscopic quasar and by WISE will appear under "Known QSOs" only. With the variables data set, the asterisked number is the total number of sources examined but the number in parentheses is the number of objects with AGN-consistent variability and WISE colours.} 
  \label{datasets}
  \begin{tabular}{llcc}
  \hline
  Sample & Number of unique & Number of AGN & Number of candidates excluded \\
  &  sources & flare candidates & due to more than one significant flare \\
  \hline
Known QSOs (A) & 321,535 & 13 & 65\\
XDQSO (B) & 324,338 & 26 & 32\\
WISE (C) & 233,373 & 10 & 27\\
Transients (D) & 2,965 & - & -\\
Variables (E) & 1,290,055* & & \\
 & (27,003) &  2 & 4\\
\hline
Total & 2,127,266* &  51 & 128 \\ 
\hline 
\end{tabular}
\end{table*}

\subsubsection{Spectroscopically-confirmed quasars}

The Million Quasars (MQ) catalogue\footnote{http://quasars.org/milliquas.htm} v3.7 contains all spectroscopically confirmed type 1 QSOs (309,525), AGN (21,728) and BL Lacs (1,573) in the literature up to 2013 November 26 and formed the basis for the results of \cite{graham15b}. We have extended this with an additional 297,301 spectroscopically identified quasars in the SDSS Data Release 12 \citep{paris15}. We crossmatched this combined quasar list against the CRTS data set with a 3\arcsec \, matching radius and find that 334,402 confirmed quasars are detected by CRTS. Of these, 12,867 do not have enough observations $(n < 5)$ for any peak to be identified, leaving a data set of 321,535 quasars.

\subsubsection{The Extreme Deconvolution Quasar Sample (XDQSO)}

\cite{dipompeo15} have combined (forced) {\em WISE} $W1$ and $W2$ photometry with SDSS colours using extreme deconvolution to produce a probabilistic catalog of 5,537,436 quasar candidates in SDSS DR8. Of these, 1,730,760 have a corresponding match in CRTS. We select those candidates with $p_{QSO} > 0.99999$ which gives 589,350 sources, of which 425,767 are not previously known quasars and have $n > 5$. We note that of the 264,821 confirmed (from MQ) quasars in XDQSO, 128,104 (48.4\%) have $p_{QSO} > 0.99999$ and 15\% have $p_{QSO} < 0.95$. Each quasar candidate in the catalog has also been assigned a photometric redshift which we use when required for cosmological calculations.

\subsubsection{WISE-selected AGN}

As part of defining a joint variability and colour-selected quasar catalog from CRTS data (Graham et al., in prep.), we have identified 233,373 WISE AGN on the ecliptic. These have W2 < 15 and W1 - W2 > 0.8, $ | b | > 10$, and $-15 \le \delta \le 15$ \citep{stern12, assef13}. They also do not appear in the known quasar or XDQSO samples (although note that with these selection criteria there are 35,731 duplicates with MQ and 109,186 duplicates with XDQSO).

\subsubsection{Transients}
To date, CRTS has detected 13,149 optical transients (see http://crts.caltech.edu for details). Many of these are not associated with any previously detected source, indicating that these sources were below the survey detection limit in their (more) quiescent state. 3,628 of these show  an apparent aperiodic variability that shows increasing amplitude on long timescales whilst lacking any obvious short timescale outbursts or rapid variations. This variability is consistent with being an AGN and ancillary data, such as colour, spectra, coincidence with a radio source, etc., is used when available to support the identification. None of these sources have been previously classified as an AGN. Given the nature of their initial detection, we have also included these in our candidate list.

\subsubsection{Variables}
CRTS is the basis for many studies of variability in astronomical populations and in an initial characterization of general source variability in CRTS, we identified a set of 1 million objects with Stetson J/K values above magnitude-dependent fiducial values in the local field (see \cite{drake14} for more details). Although these sources remain largely unclassified, we are including them in this analysis since they can be used to estimate flaring statistics for a general variable population in addition to providing more AGN candidates. We have determined QSO variability statistics and WISE colours for these and select 27,003 candidates which are not part of any of the other data sets described here.

\section{Identifying major flaring activity}

\cite{lawrence16} defines an AGN as a {\em slow blue hypervariable} if it has brightened by an order of magnitude $(|\Delta g| > 1.5)$ over the course of a decade in a smooth fashion and is now mostly fading (but may also still be increasing in flux). The strength of its Mg II emission may be weaker as well than what would be expected given its luminosity. We are interested in finding any AGN source with outburst activity that can be characterized as statistically significant. This relates not only to the strength of the activity but also its duration and morphology. Because of the novelty of this field, the relevant distributions are a combination of theoretical and phenomenological.

\subsection{The amplitude of variability}

Although a number of examples of extreme quasar variability are known, particularly in the blazar population,
to date the statistics of extreme variability have only been marginally constrained. Assuming a CAR(1)/DRW model for quasar variability in a sample of 33,881 quasars with at least two epochs of SDSS or POSS photometry, \cite{macleod12} found that the distribution of magnitude difference for a given time lag is exponential. This is a cumulative effect of averaging over a range of different characteristic timescales, $\tau$, and variability amplitudes, $SF_\infty$. From a joint SDSS-PS1 analysis, \cite{lawrence16} claims that somewhere in the range 1 in 1,000 to 10,000 AGN show extreme variability ($| \Delta g |> 1.5$) over the period of a decade.

\begin{figure}
\centering
\caption{Joint probability distribution of magnitude difference and time lag from the ensemble of known quasars. The sampling effects of annual cycles can be clearly seen.}
\label{dmmed}
\includegraphics[width = 3.3in] {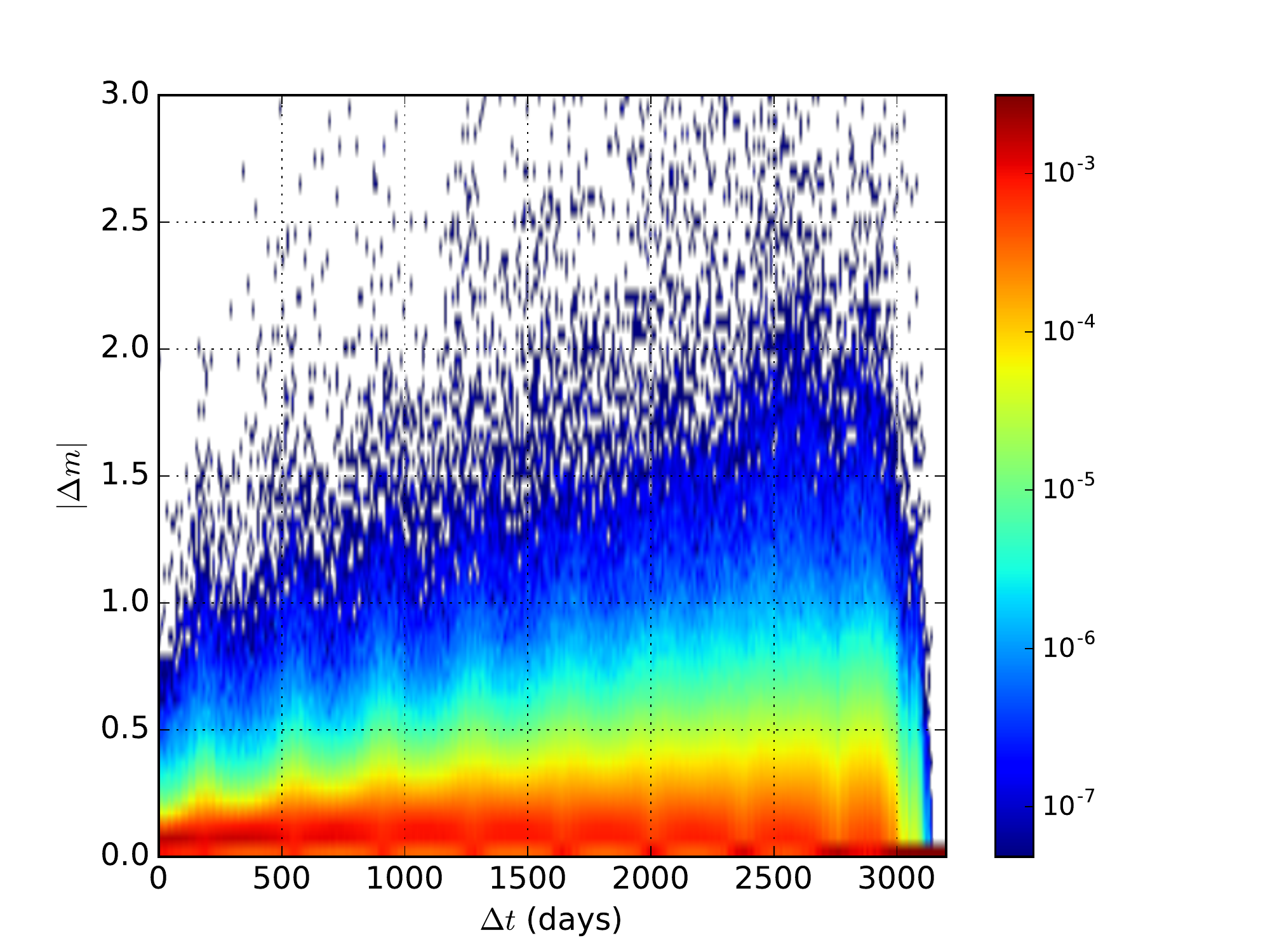}
\end{figure}

\begin{figure}
\centering
\caption{Cumulative absolute magnitude difference seen in CRTS data from different time lags compared to $r$-band predicted values for PTF for r < 20.6 from a DRW model of quasar variability (MacLeod et al. 2012). For the purposes of this analysis, CRTS and PTF are comparable - a $r < 20.6$ limit is equivalent to $V_{CRTS} < 20$.}
\label{cummagdiff}
\includegraphics[width = 3.3in] {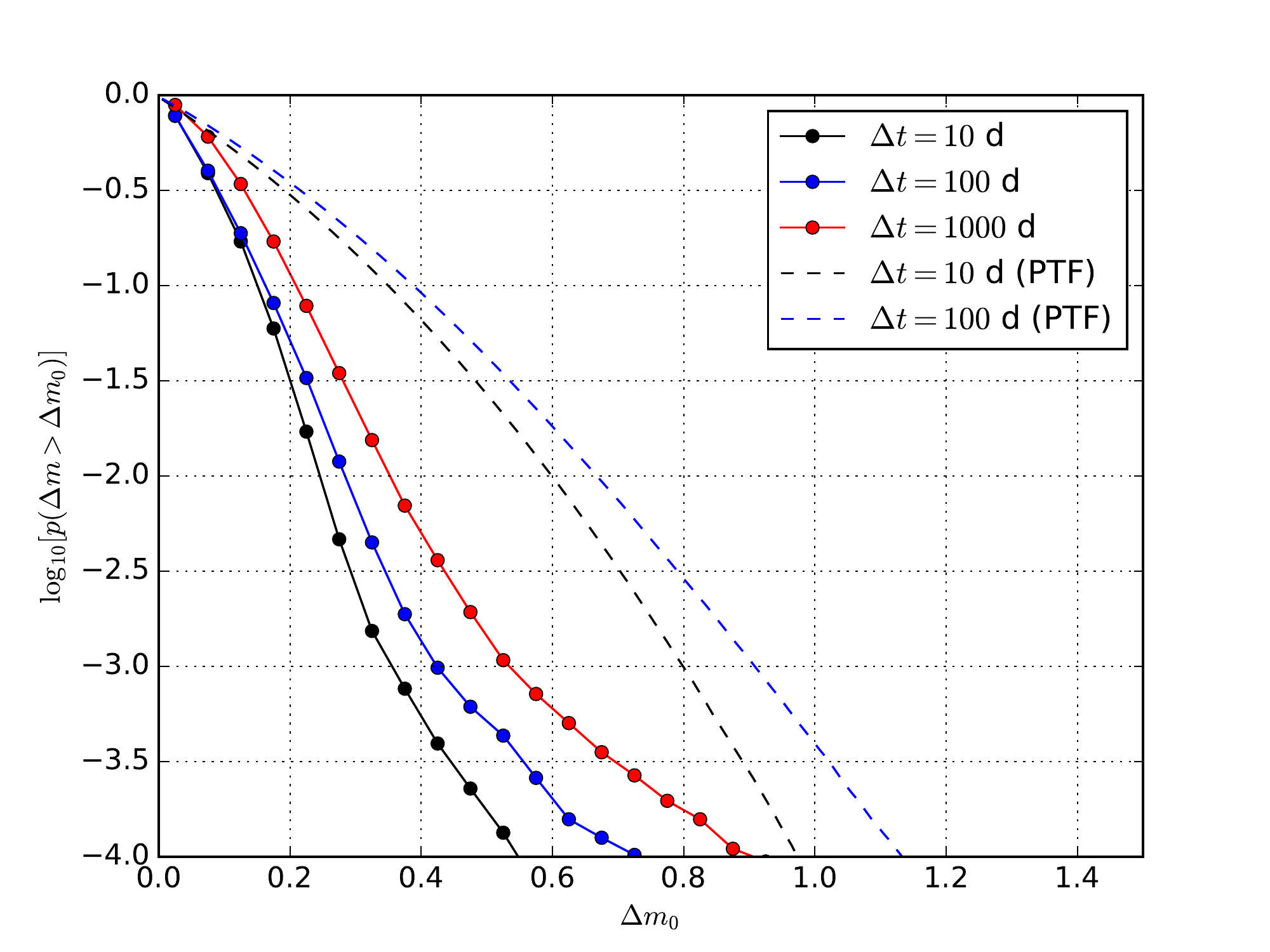}
\end{figure}

The sampling and time coverage of CRTS data allows us to validate these claims. For each quasar in the known data set (A), we have calculated the median absolute magnitude difference as a function of time lag, $med(|\Delta m (t)|)$, spanning a range of 10 to 3,200 days in bins of $dt = 10$ days. Fig.~\ref{dmmed} shows the ensemble joint probability distribution, from which we can determine the expected number of quasars with variability $| \Delta m | > \Delta m_0$ over a timescale $t \ge t_0$.  Fig.~\ref{cummagdiff} shows the observed cumulative magnitude differences against the predicted behavior for three different time lags. Previous structure function-based analyses (de Vries et al. 2005, \cite{macleod12, morganson14, kozlowski16} have shown that the amplitude of AGN variability increases with longer timescales and the behavior we see in our data set is consistent with this.

\begin{figure}
\centering
\caption{Marginal probability distribution for median absolute magnitude differences $|\Delta m|$ over 3,200 days.}
\label{probdm}
\includegraphics[width = 3.3in] {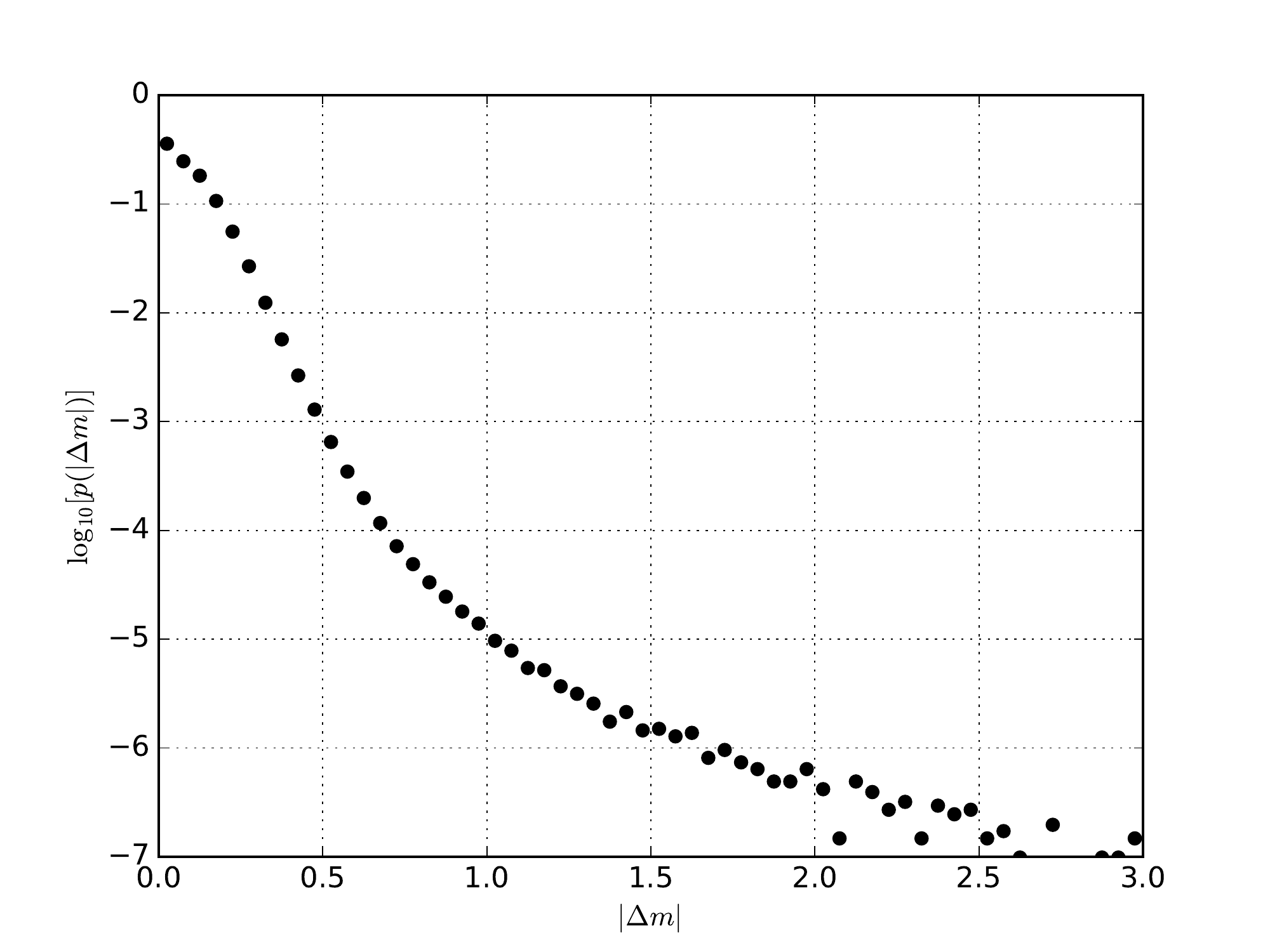}
\end{figure}

Fig.~\ref{probdm} shows the marginal probability distribution for $ | \Delta m | $.  This suggests that extreme variability is an order of magnitude rarer than previous claims and that a limit of $| \Delta m | > 1.0$ of $\sim$10 years $p(|\Delta m| > 1) \sim 10^{-3}$ is sufficient to define significant flaring activity.

\subsection{Selection criteria}
\label{select}

We model the median activity of a source over time with a linear fit to its light curve. We derive the slope and best offset using the Thiel-Sen estimator (the median slope between all pairs of points). Candidate flares are identified as contiguous sets of points brighter than the median level and each is characterized in terms of its temporal span ($tspan)$, median (and peak) amplitude $(amp)$ above the median level (to help identify false positives resulting from poor quality photometry), and number of contributing points $(npts)$. Since we are interested in the joint distribution of these three features, we found that we could also represent this through a single parameter, defined as the logarithm of the geometric mean of the three features:

\[ pksig = \frac{1}{3}[\log_{10} (amp_{norm}) + \log_{10}(npts) + \log_{10}(tspan)] \]

\noindent
Note that $amp_{norm}$ is the normalized median amplitude (see below). A typical given source will contain several such flares and we use the median absolute deviation of its flare amplitudes to define a baseline level of flare activity for the source against which to identify significant flares. 

\begin{figure}
\centering
\caption{Median flare amplitude as a function of magnitude for 87,017 spectroscopically confirmed known quasars with some degree of flaring (timespan $> 100$ days and more than 10 observations in the flare). The black dashed line indicates the median absolute deviation as a function of magnitude for 72,634 standard stars in Stripe 82 which is used to normalize the variability scale.}
\label{ampmag}
\includegraphics[width = 3.3in] {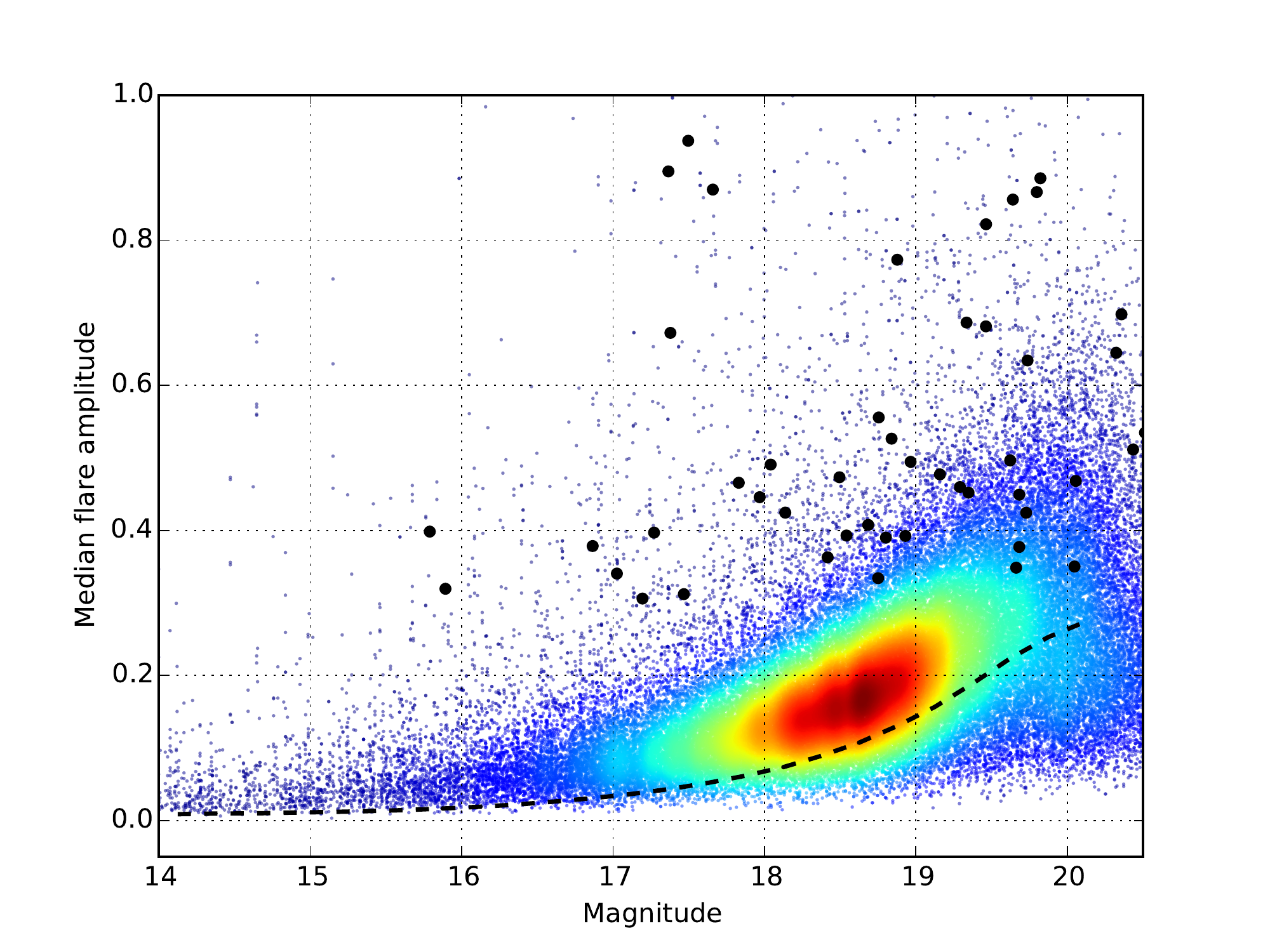}
\end{figure}

The scale of variability is a function of magnitude: fainter objects show larger median absolute deviations (MAD) as there is a larger noise contribution for low S/N (faint) sources (see Fig.~\ref{ampmag}). Using CRTS data for 72,634 standard stars in the Stripe 82 region \citep{ivezic07} with a magnitude range of $V = 14$ to 20.5, we have derived a normalization based on the median MAD value for a given magnitude to ensure that objects with equivalent variability strength (irrespective of magnitude) can be compared. 

\begin{figure*}
\centering
\caption{Sample Weibull distributions. The left plot shows varying the scale for fixed shape $(a = 2)$ and the right changing the shape for fixed scale $(s = 600)$.}
\label{weibull}
\includegraphics[width = 6.85in]{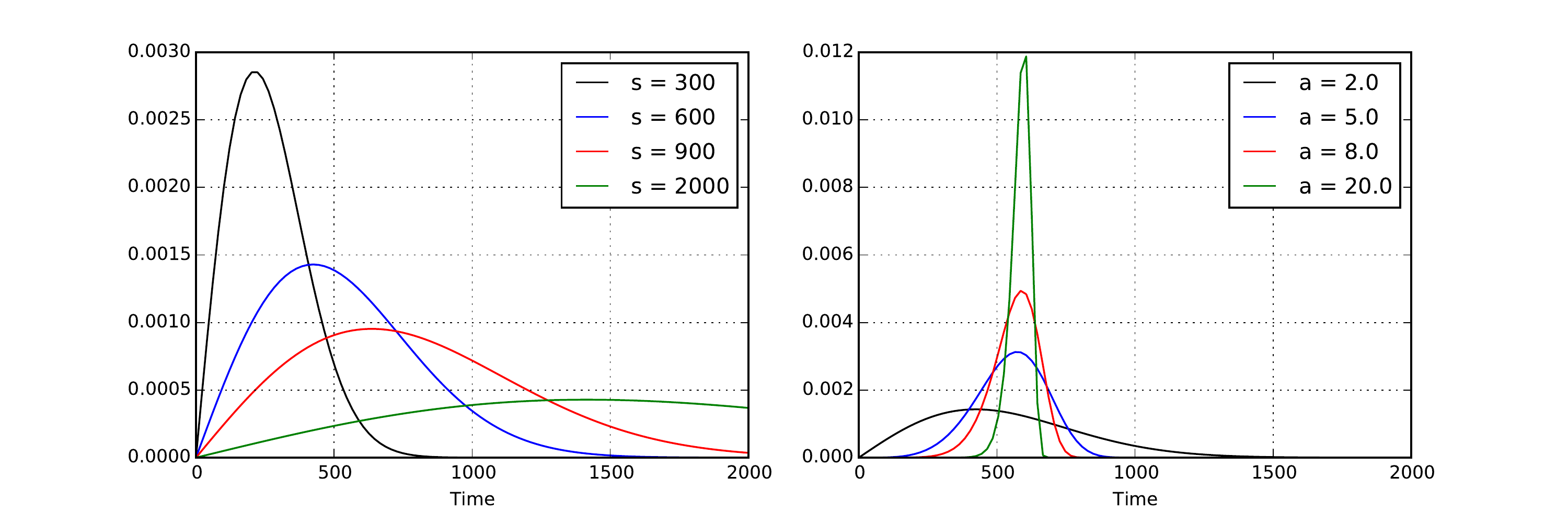}
\end{figure*}

We also characterize any flares in terms of shape parameters using the (translated) Weibull distribution. This has been shown to be a convenient function for empirically fitting the shapes of flares \citep{huenemoerder10}. The probability distribution is:

\[ f(p; a, s) = \left( \frac{a}{s} \right) p^{a-1}e^{-p^{a}} \]
\[ p = (t - t_0) / s\]

\noindent
in which $a$ is a shape parameter $(a > 0)$, the scale (or width) is specified by $s (s > 0)$, and the offset (location) is given by $t_0 (t_0 \ge 0)$; the independent coordinate is $t (t \ge t_0)$. For fixed $a$, increasing values of $s$ stretch out/broaden the function and for fixed $s$, increasing values of $a$ sharpen the peak (see Fig.~\ref{weibull}). We fit $Af(p;a,s) + R_0$ (relative to the linear fit representing the median source activity), where $A$ is an amplitude and $R_0$ is the baseline magnitude. Note that this is just a location-scale transformed version of the Weibull distribution. 

Given the apparent rarity of significant flares, we expect such objects to be population outliers in the parameter space defined by these characterizing features. From the two examples in the literature \citep{meusinger10, drake11}, we might expect the outburst to last at least 300 days (in the observed frame) and have a peak of at least 1 mag above the normal level of activity which is consistent with our predictions in the last section. We note that non-type IIn supernovae have typical decay times of between 20 and 150 days and so most of these phenomena would be excluded. We use a lower cutoff of 30 observations in the light curve to ensure that the flares are sufficiently well-sampled.

We expect to see only one significant flare over the time covered by the light curve and exclude those with more -- these are typically cataclysmic variables or carbon stars, a consequence of including quasar candidates in our sample. We also want to exclude blazars as we are specifically interested in individual flaring events from a (semi-)quiescent state rather than general continuous large amplitude ($> 1$ mag) flaring activity. One possible source of contamination is stray light (e.g., diffraction spikes, reflections) from a nearby bright source, low surface brightness galaxy or genuine blend. We applied the same criteria as described in \cite{graham15b} to identify such sources and exclude them from further consideration.

\subsection{Mock data}

\begin{figure}
\centering
\caption{A mock light curve for a damped random walk process.}
\label{mockqso}
\includegraphics[width=3.3in]{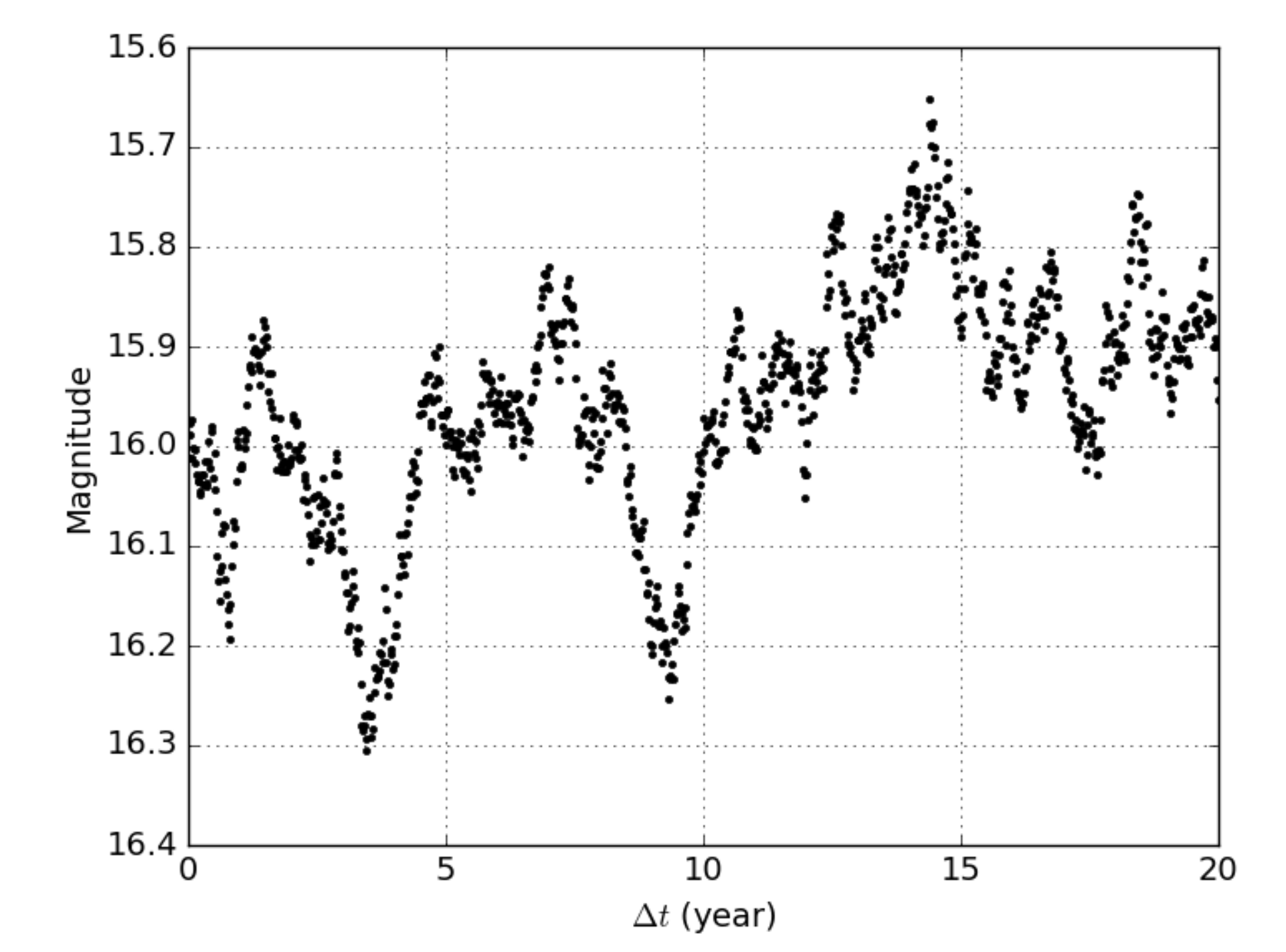}
\end{figure}

Autoregressive processes, such as those used to describe quasar variability \citep{kelly14}, have correlated (red) noise (characterized by a power spectrum of the form $P(f) \propto \nu^{-2}$) which can introduce features in their time series, such as dips and humps (see Fig.~\ref{mockqso}). To ensure that the features we are identifying are not just noise artifacts (note that these are intrinsic to the source and not associated with any form of measurement noise), we can determine the expected distribution of noise-related features in terms of the characterizing measures we are employing to describe the real flares -- median amplitude and time span -- from simulated time series generated by a particular stochastic model. As in previous analyses \citep{graham14, graham15b}, we simulate quasar variability via a DRW process characterized by a timescale $\tau$ and an amplitude $\sigma^2$. A (zero centred) data point  $m_{i+1}$ at time $t_{i+1}$ is given by:

\[ m_{i+1} = m_i e^{-\Delta t / \tau} + G \left[ \sigma^2 \left(1 - e^{-2 \Delta t / \tau} \right) \right] \]

\noindent
where $G(s^2)$ is a Gaussian deviate with variance $s^2$ and $\Delta t = t_{i + 1}  - t_i$. A second Gaussian deviate is added to each value to represent observational noise: $y_i = m_i + G(n_i^2)$, where $n_i$ is the error at time $t_i$ - this ensures heteroscedastic errors as in the real light curves. Previously we have drawn $\tau$ and $\sigma^2$ from the rest-frame fitting functions determined by \cite{macleod10} but we have now calculated $(\tau, \sigma^2)$ for all known quasars via Gaussian process regression. We evaluated the mean and covariance of these in $\Delta m = 0.5$ magnitude bins and now draw a magnitude-dependent random $(\tau, \sigma^2)$ from the joint Gaussian with the appropriate mean and covariance (see Table~\ref{mockpars}).

\begin{table}
  \centering
  \caption{The magnitude-binned means and covariances of the DRW parameter distributions.}
  \label{mockpars}
  \begin{tabular}{lllllll}
  \hline
 Magnitude & $<\!\!\tau\!\!>$ & $<\!\!\sigma^2\!\!>$ & $C_{00}$ & $C_{01}$ & $C_{10}$ & $C_{11}$ \\
  \hline
14.25 & 1.966 & -2.422 & 1.142 & 0.173 & 0.173 & 0.379\\
14.75 & 2.186 & -2.531 & 0.772 & 0.105 & 0.105 & 0.476\\
15.25 & 2.160 & -2.494 & 1.046 & 0.343 & 0.343 & 0.469\\
15.75 & 2.375 & -2.314 & 0.768 & 0.208 & 0.208 & 0.407\\
16.25 & 2.535 & -2.339 & 0.636 & 0.197 & 0.197 & 0.312\\
16.75 & 2.613 & -2.225 & 0.561 & 0.191 & 0.191 & 0.266\\
17.25 & 2.753 & -2.137 & 0.299 & 0.116 & 0.116 & 0.218\\
17.75 & 2.798 & -2.047 & 0.252 & 0.120 & 0.120 & 0.201\\
18.25 & 2.794 & -2.018 & 0.228 & 0.106 & 0.106 & 0.172\\
18.75 & 2.704 & -1.938 & 0.266 & 0.108 & 0.108 & 0.181\\
19.25 & 2.580 & -1.758 & 0.306 & 0.098 & 0.098 & 0.160\\
19.75 & 2.409 & -1.611 & 0.406 & 0.074 & 0.074 & 0.115\\
20.25 & 2.221 & -1.602 & 0.507 & 0.033 & 0.033 & 0.080\\
\hline
\end{tabular}
\end{table}

\section{Results}
\label{results}

\begin{figure}
\centering
\caption{Frequency distribution for flare timespans and median amplitudes. The apparent periodicity in the flare timespan distribution is due to the sampling effects of annual cycles as in Fig.~\ref{dmmed}.}
\label{flares}
\includegraphics[width = 3.3in] {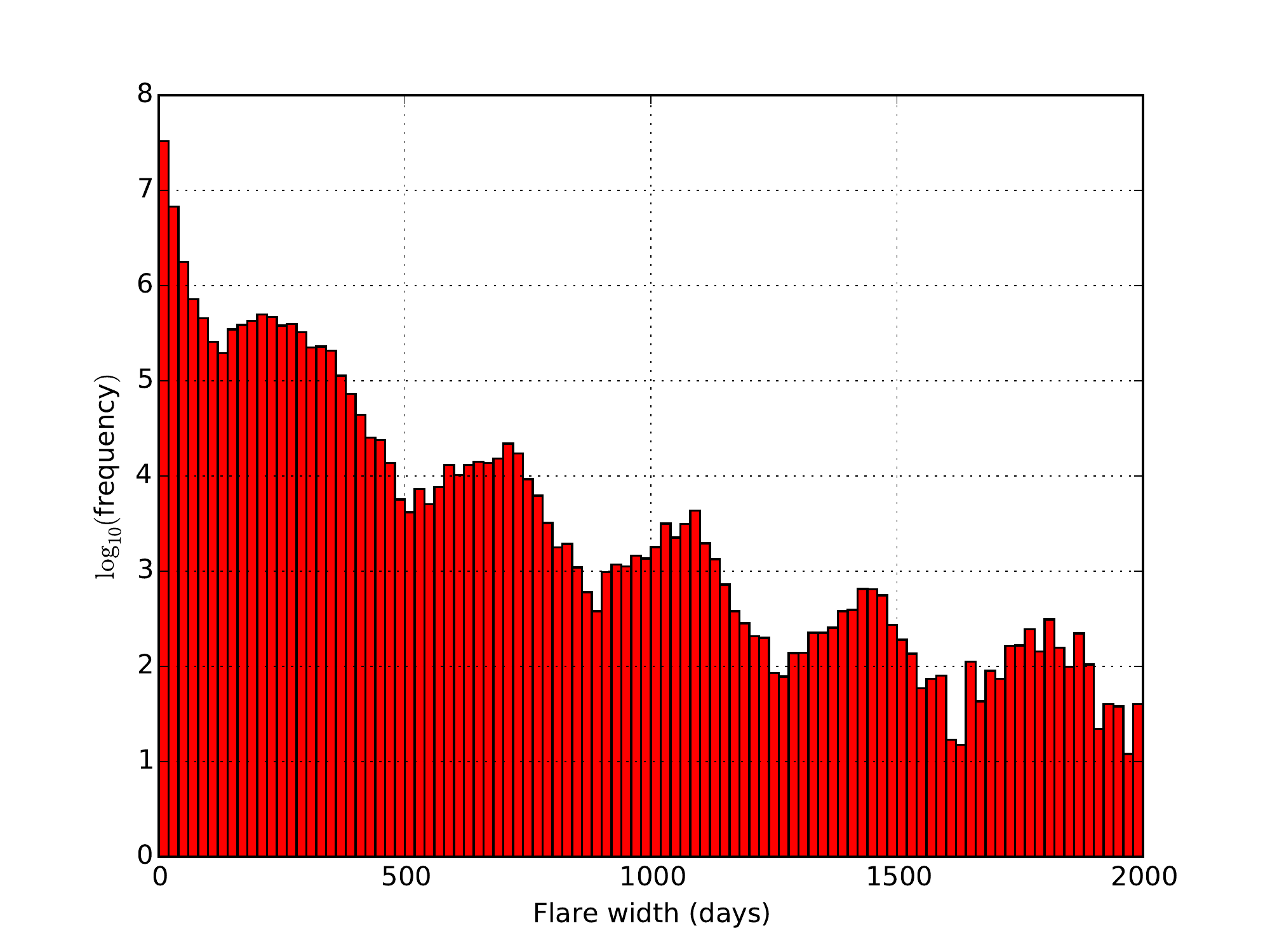}
\includegraphics[width = 3.3in] {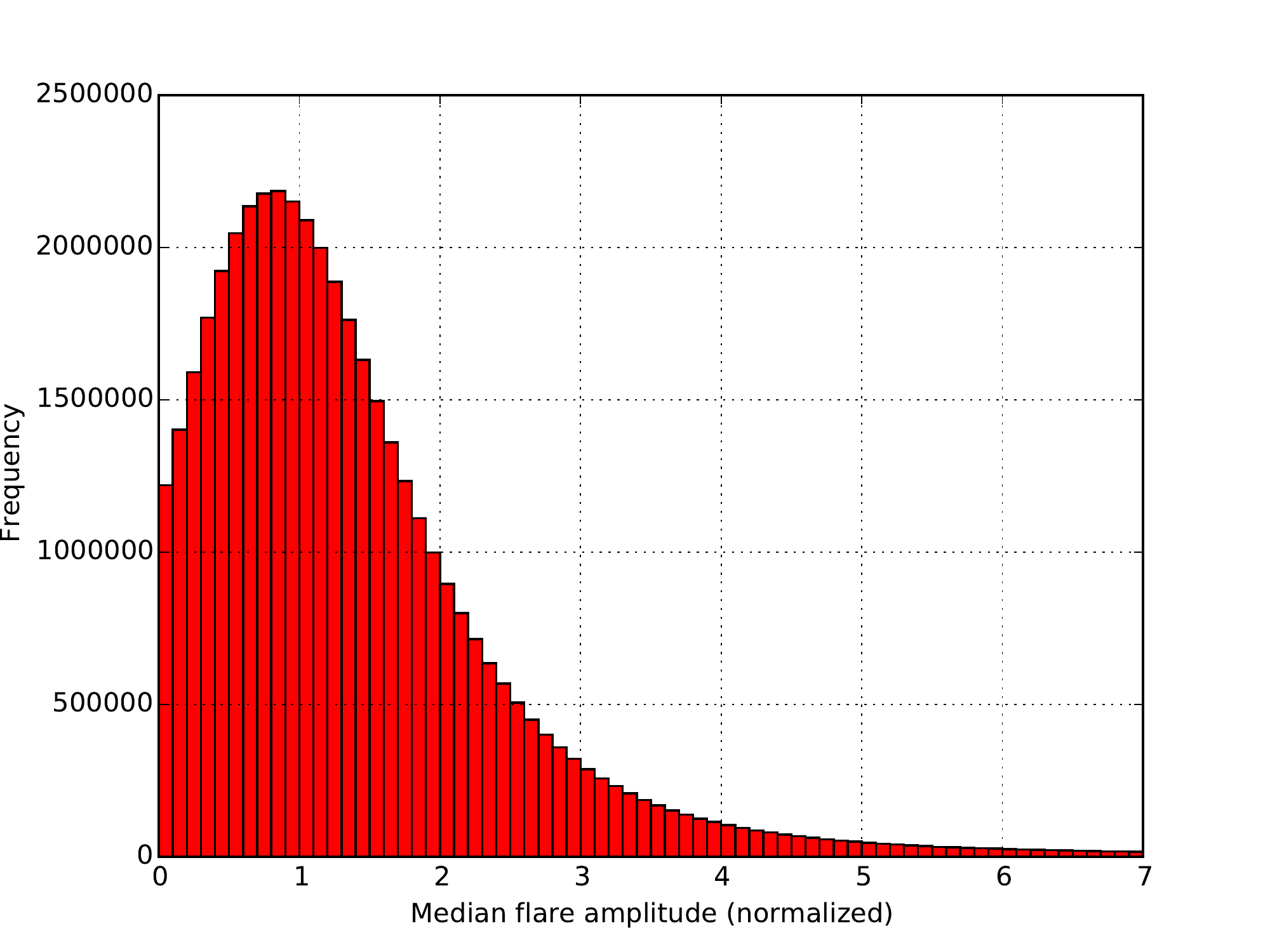}
\end{figure}

We applied our flare identification algorithm to 2,127,266 CRTS light curves and found 91,321,768 candidate flares (see Fig.~\ref{flares}). An initial selection is provided by selecting those flares with a timespan longer than 300 days, a normalized median amplitude greater than a factor of 2.5 and sampled by 30 or more observations. This produces 19,150 flares from 14,592 distinct sources (known AGN or AGN candidates). Removing those associated with nearby bright stars or blended sources reduces this to 13,527. 
We have also ignored activity from 529 known blazars (using the class designation in MQ and the BZCAT v5.0 catalog of blazars \citep{massaro15}). For comparison, we find 1,602 from a simulated realization of the known quasars sample containing 321,535 sources and the same selection criteria. This suggests that the number of real flares is more than expected.

\begin{figure}
\centering
\caption{Distribution of peak flare amplitude against the flare significance. The small points are colour coded according to the local density of points. The dotted red line indicates the 0.5 cutoff value we use and the larger black dots are the identified flare candidates.}
\label{pkmax}
\includegraphics[width = 3.3in] {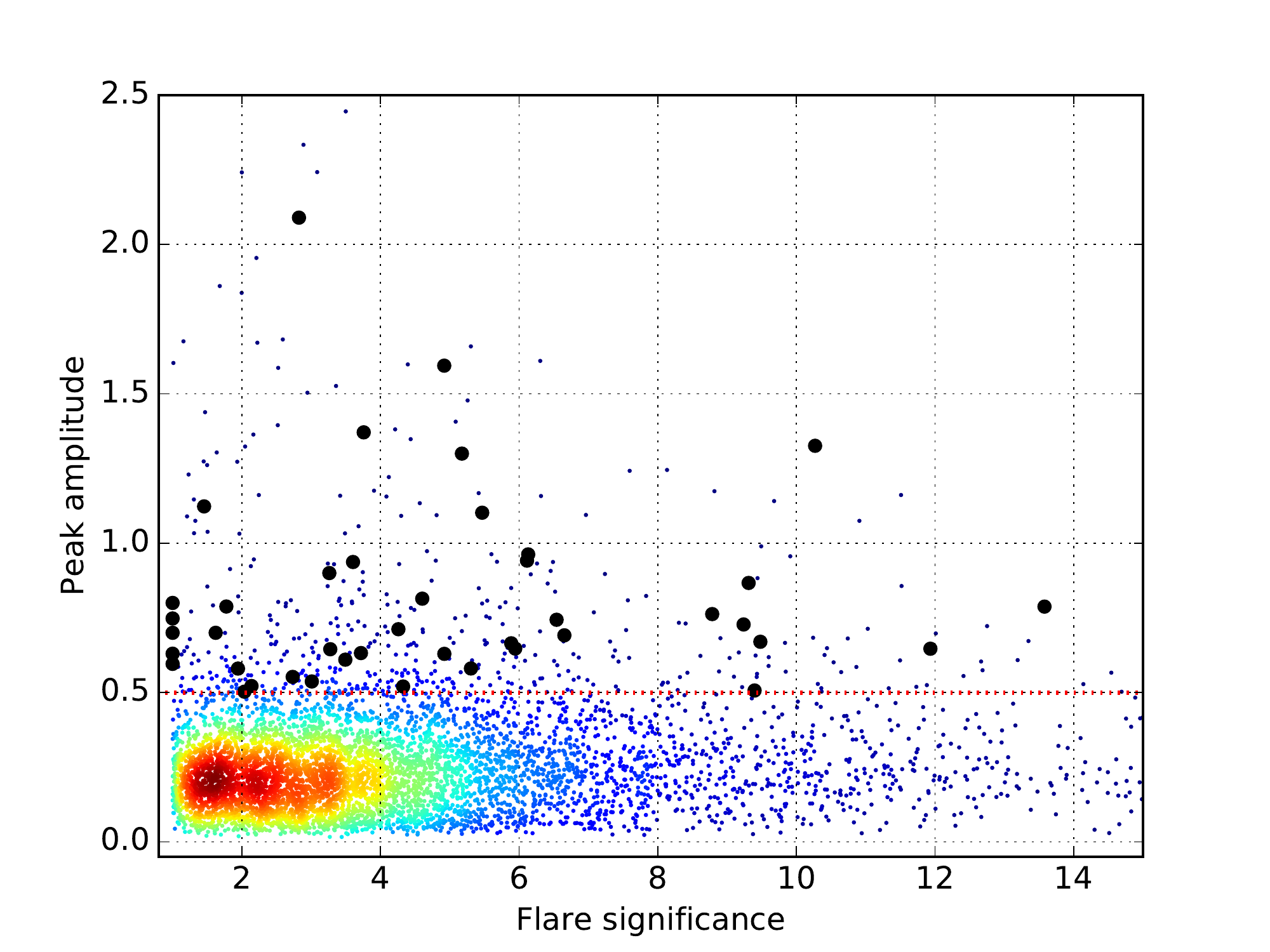}
\end{figure}

We are interested in those sources where the flaring represents a noticeable change from a lower or quiescent state over the timespan of the light curve, i.e. where the flaring activity is significant relative to the baseline activity of the source. From the distribution of peak amplitude against the significance of the flare: 

\[sig = \frac{pksig - pksig_{med}}{ mad(pksig)} \] 

\noindent
we identify a set of 585 candidates (see Fig.~\ref{pkmax}) with peak amplitude $> 0.5$ mag (154 sources from the simulated data pass this criterion). This is more inclusive than the $\Delta m = 1$ limit we argue for in Sec.~\ref{select} but captures the significant large amplitude outliers from the sample distribution. We also note in Table~\ref{datasets} the number of candidates that were excluded due to the presence of more than one significant flare in the light curve.

The flaring could still be the result of correlated noise (regular quasar variation) rather than a specific physical mechanism. For each candidate, we therefore construct a comparison time series with the primary flare removed. We describe both time series (with and without the flare) as a DRW process\footnote{This is a purely statistical description of the variability and makes no inferences about the physical processes contributing to variability.} via Gaussian process regression and incorporating heteroscedastic errors (using the Python code GPy\footnote{http://gitlab.com/GPy}). These models are parameterized by a characteristic timescale $\tau$ and a variance $\sigma^2$. If the flaring activity is consistent with the general variability of the quasar (arising from correlated noise in a DRW) then the sets of $(\tau, \sigma^2)$ values for the two light curves should agree within the confidence limits on the parameters. However, differing values indicate that the flaring is incompatible with such a model.

\begin{figure}
\centering
\caption{Distribution of DRW parameter differences between the time series with and without the primary flare. 
The black dotted line indicates the contour level used to identify outliers (flare candidates). Two outliers lie outside the bounds of the plot. The blue star denotes the position of Sharov 21 with this analysis. }
\label{drwdiff}
\includegraphics[width = 3.5in]{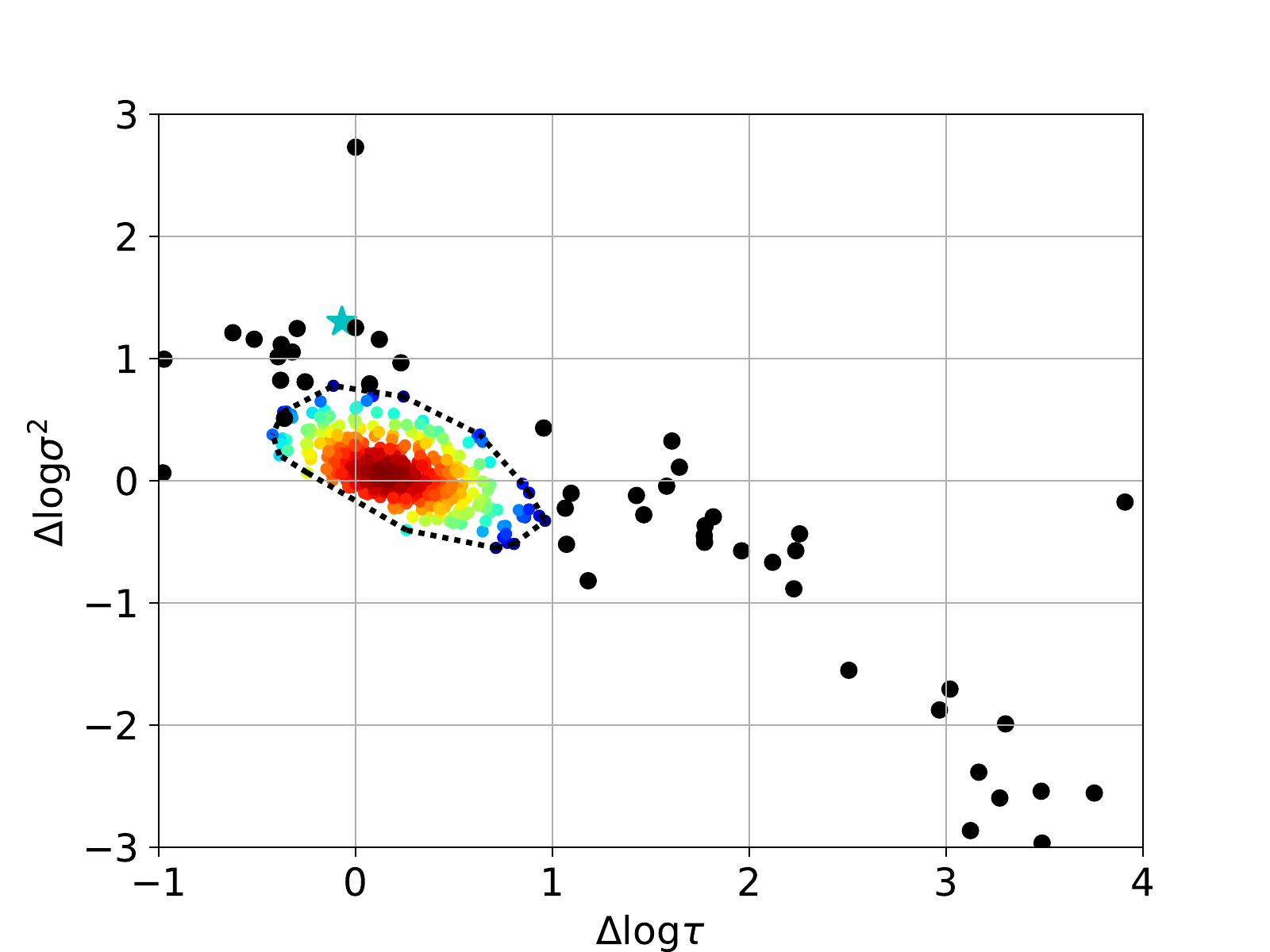}
\end{figure}

Fig.~\ref{drwdiff} shows the distribution of the parameter differences between the two time series. Most of the sources have compatible DRW descriptions with or without the flare (clustered around the origin); however, a subset of 48 do not and we consider these to be major flare candidates. We also note that there are three objects where there is no discernible lower state, i.e., the ``flare" represents the whole light curve. Determining a parameter difference is therefore not possible with these but the parameter values from describing these with a DRW are sufficiently different from the general population that we regard them as a separate set of ``superflare" candidates (although changing-look candidates are also on the top right of the DRW parameter $\tau$--$\sigma^2$ distribution (Graham et al., in prep.).

Table~\ref{candidates} lists all the flare candidates and their light curves are shown in Fig.~\ref{lightcurves}. Where possible, we have obtained spectroscopic redshifts for candidates without existing spectra using the Palomar 200'' and Keck telescopes (see appendix for more details). Photometric redshifts are used for those remaining candidates without spectra (which are typically outside the SDSS footprint). These are taken from the XDQSO catalog or in those cases where one is not available, calculated using the XDQSO IDL code\footnote{http://xdqso.readthedocs.org/} with SDSS magnitudes and forced WISE photometry from \cite{lang14}. We fit each flare candidate with a Weibull distribution (as described in Sec.~3.2) to allow further characterization of the phenomena that we are detecting and the parameters are reported in the table.

We also estimate the total energy output by each flare. The absolute magnitude of a source is given by: 

\[ M_V = m_V - A_V - DM - K_V \]

\noindent
where $A_V$ is the extinction, $DM$ is the distance modulus and $K_V$ is the K-correction. We obtain\footnote{http://irsa.ipac.caltech.edu/applications/DUST/} extinction values at the source position from the \cite{schlafly11} recalibration of the \cite{schlegel98} reddening maps. We assume a K-correction of: $K = -2.5 (\alpha + 1) \log (1 + z)$ for a power law SED of $F_{\nu} \propto \nu^\alpha$ with $\alpha = -0.5$. The bolometric luminosity in band $X$ can be defined as:

\[ L_{bol,X} = b_X L_{\odot,X}10^{(M_{\odot,X} - M_X) / 2.5} \]

\noindent
where the solar constants for filter $V$ are $M_{\odot,V} = 4.83$ and $L_{\odot,V} = 4.64 \times 10^{32}$ erg s$^{-1}$ and $b_X$ is the bolometric correction. We report the total integrated bolometric luminosity without bolometric correction for each source in Table~\ref{candidates}. As a check, \cite{drake11} give a value of $8.5 \times 10^{50}$ erg for this quantity for the source J102913+404420 which compares well with our estimate of $8.63 \times 10^{50}$ erg. They also determine a mean bolometric correction of $\bar{b}_V \sim 15$ giving an integrated bolometric luminosity of $\sim 1.3 \times 10^{52}$ erg. Although the quantity can be a source- and time-dependent quantity, particularly during the flaring activity, we will adopt a canonical value of $b_V = 10$ for the candidates here.

\begin{figure}
\centering
\caption{Frequency distribution for total energy output by the flare.}
\label{erghist}
\includegraphics[width=3.3in]{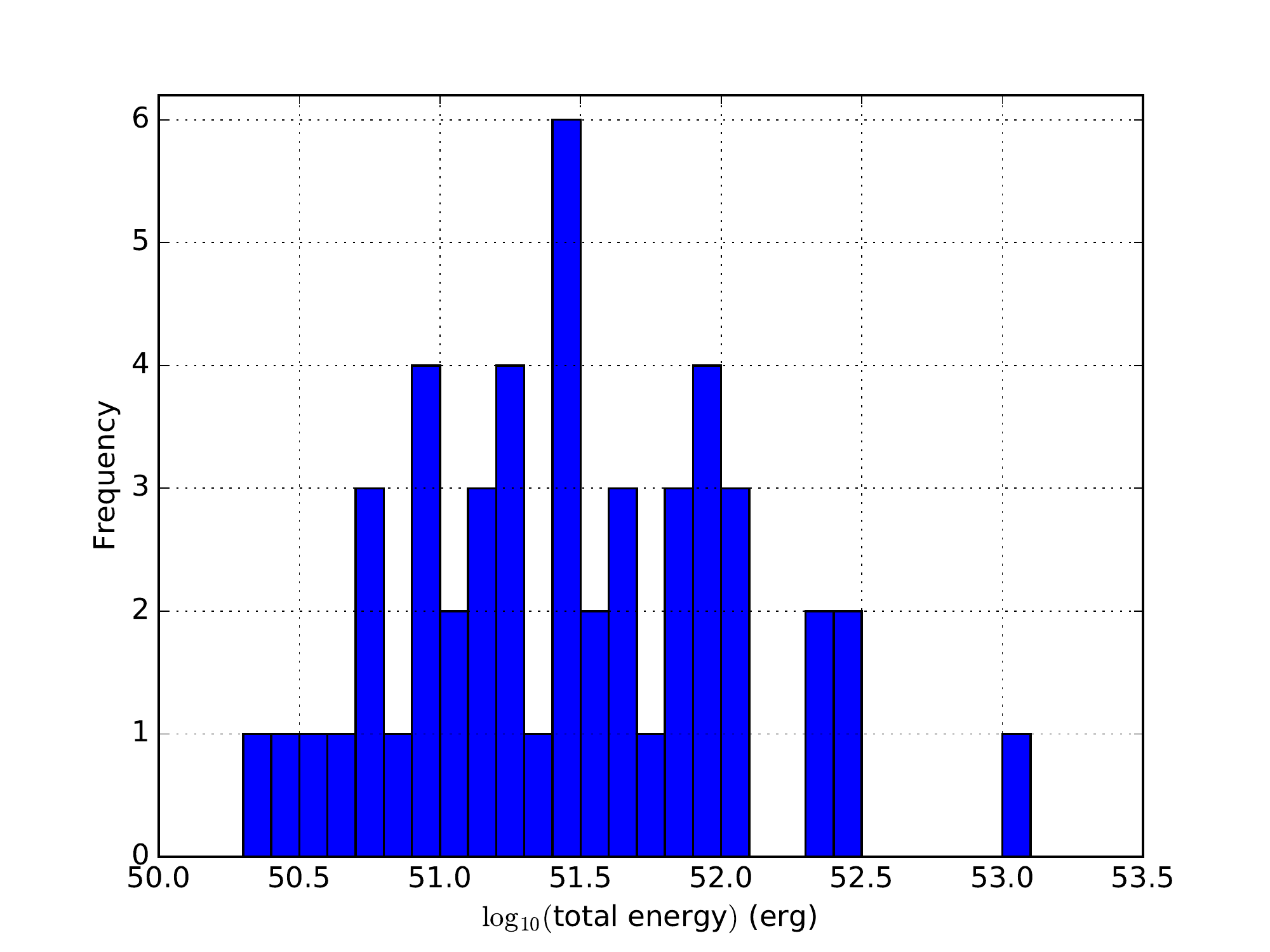}
\end{figure}

\section{Discussion}

In this section, we consider possible physical explanations for the extreme AGN variability we have detected.

\subsection{Microlensing}

From 3 years of PS1 data, L16 identified 49 AGN, most previously unknown, with variability ostensibly similar to what we have reported: smooth order of magnitude outbursts over several years. They also identified an additional 15 AGN in SDSS Stripe 82 data with $\Delta g > 1.5$ over a decadal baseline. They propose that microlensing by the close passage of a single star in an intervening galaxy, or a caustic caused by a small number of stars, is the most likely explanation for such rare temporary large amplitude events. Furthermore, a canonical model with a source at $z = 1$ and a $1 M_\odot$ mass star in motion within a galaxy at $z = 0.25$ with a transverse velocity of 300 km s$^{-1}$ predicts a characteristic timescale and event rate within an order of magnitude of that seen. Sharov 21 \citep{meusinger10} is considered a canonical example in the existing literature, although it is slightly better fit by a two star lens model.

Assuming that both the source and the lens are pointlike and that their relative motion is linear, the magnification associated with microlensing can be approximated as:

\[ \mu(t) = \frac{F_{\nu,obs}(t)}{F_{\nu,gs}} = \frac{u(t)^2 + 2}{u(t)\sqrt{u(t)^2 +4}} \]

\noindent
where $u(t)$ is the angular distance between the source and the lens in units of the Einstein angle, and $F_{\nu,obs}(t)$ and $F_{\nu,gs}$ are the observed flux density at time $t$ and the mean flux density in the ground state, respectively. This produces a symmetric profile and we can use the Weibull characterization of each flare as an indicator of how likely this model seems. 

We can test whether the range of Weibull parameters that we find is consistent with that expected from a microlensed population of quasars observed with a CRTS-like survey. For a given source at redshift $z_s$, we assume a lens at half the corresponding angular diameter distance with a transverse velocity of 300 km s$^{-1}$ and mass drawn from the galactic stellar mass function of \cite{chabrier03}. We also assume a minimum impact parameter, $u_{min}$, drawn from a uniform distribution in the range $0.063$ to $0.63$, where the lower bound comes from the maximum amplitude detected ($\Delta m = 3)$ and the upper bound from the minimum amplitude required ($\Delta m = 0.5)$ to be detected by our process. We model the redshift distribution of the quasar population from that of the CRTS + XDQSO data with $mag < 19$ to allow a reasonable detection of a flare (note that the mean magnitude of the flare candidates is 19.05). We also model the time difference between successive observations of CRTS light curves to generate equivalent irregular sampling patterns over the timescale of a lensing event. We use a ground state magnitude drawn from the magnitude distribution of the same CRTS + XDQSO data set used for the redshifts. Finally, we add heteroscedastic Gaussian noise terms to all magnitudes drawn from a Gaussian with mean and standard deviation equal to that of typical measurement errors in CRTS data as a function of magnitude. 

\begin{figure*}
\centering
\caption{The distribution of the Weibull shape $(a)$ and scale $(s)$ parameters for the flare candidates (black), and simulated flares (blue) according to a single-point single lens model. The right plot shows an enlargement of the main simulated distribution together with 1$\sigma$ confidence limits on the parameters. Dashed contour lines are shown for 20\% density increments. Red stars indicate the lensing candidates.}
\label{weibullfit}
\includegraphics[width = 3.4in] {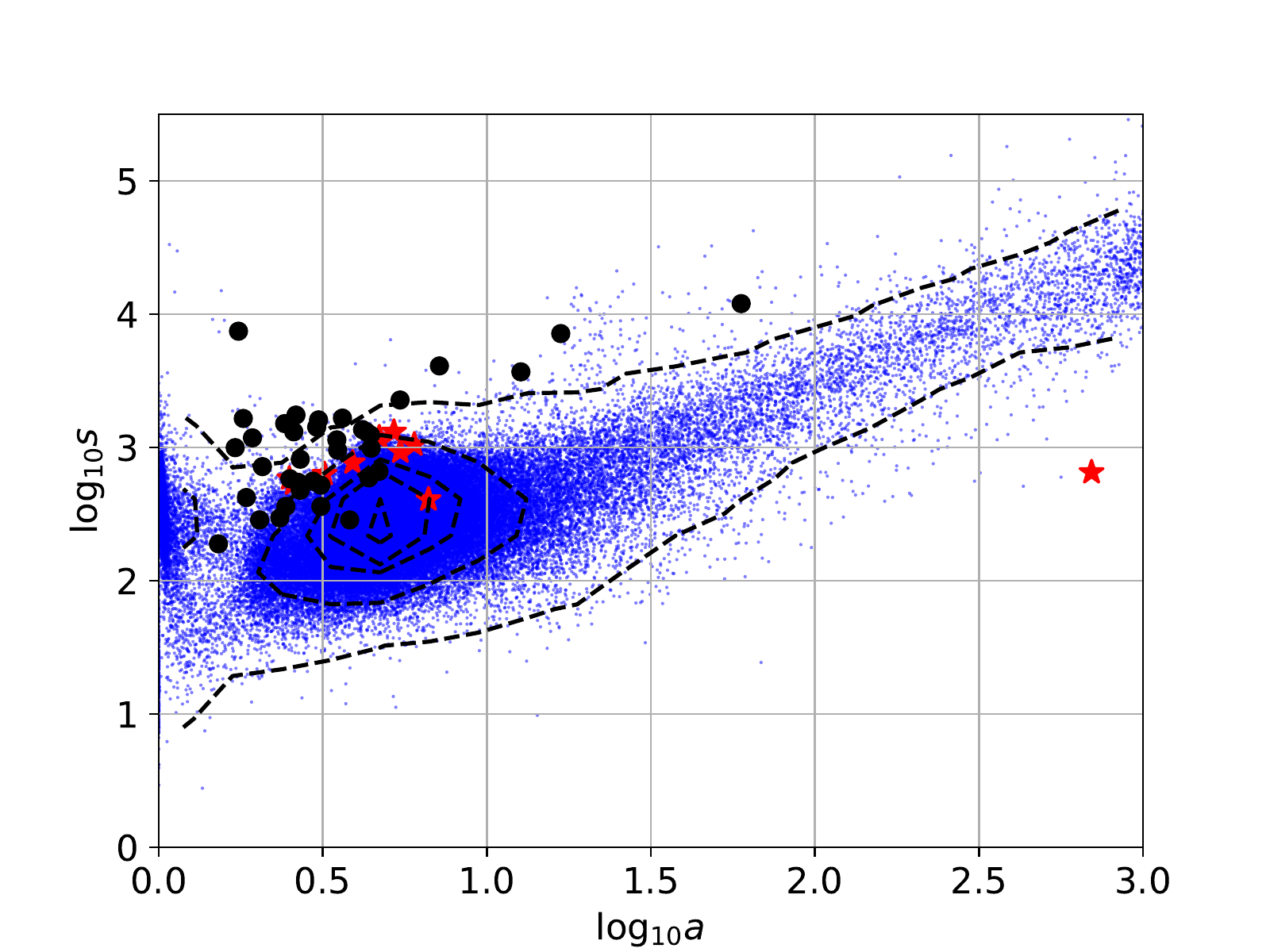}
\includegraphics[width = 3.4in] {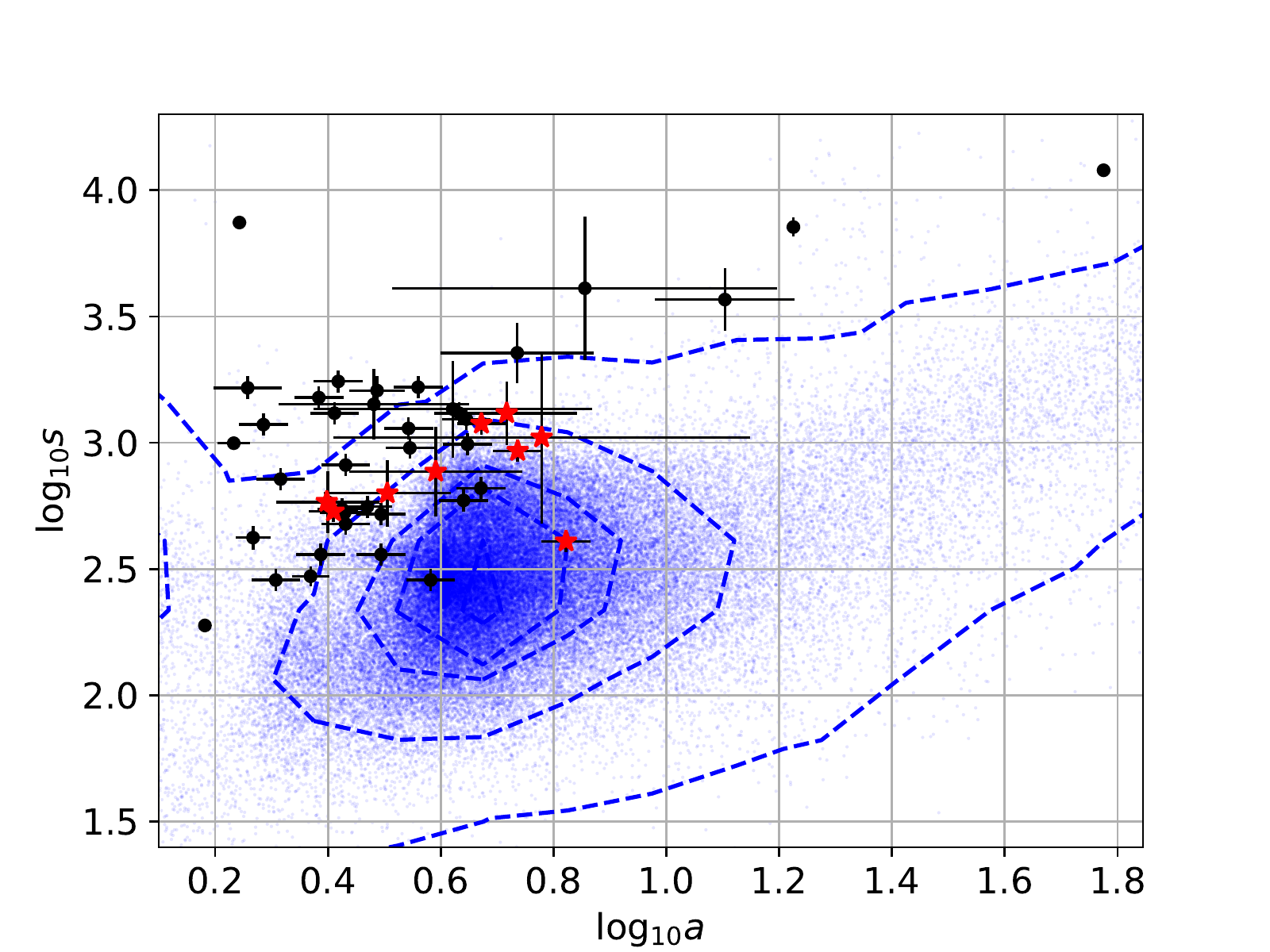}
\end{figure*}

With these priors (summarized in Table~\ref{priors}), we generated 100,000 simulated single-lens events and fit Weibull models to the resulting light curves using the same procedure as before. Fig.~\ref{weibullfit} shows the distributions of the Weibull shape ($a)$ and scale ($s$) parameters from the simulated flares as well as those from our flare candidates. It is interesting to see whether there is any relationship between the duration of a flare and its symmetry. Fig~\ref{skewdur} shows the distribution of the skewness of the Weibull fit for each flare (see Appendix~\ref{weistat} for a derivation) vs. the duration of the flare. It is clear that the symmetry of the flare is largely independent of its duration.
We have also tested simple two-lens models to add a degree of asymmetry to the simulated flare: each component is treated as a single star with multiplicative magnification \citep{meusinger10}. The resulting distribution is essentially the same as that from the single-point single lens model and so we do not consider it any further here. 

\begin{table*}
  \centering
  \caption{The prior distributions used for simulating (and fitting) single-point single/double-lens models.}
  \label{priors}
  \begin{tabular}{lll}
  \hline
  Parameter & Symbol & Distribution \\
  \hline
  Source redshift & $z_s$ &  CRTS+XDQSO with $V < 19$ \\
  Background flux & $F_{\nu,gs}$ & CRTS + XDQSO with $V < 19$ \\ 
  Lens redshift & $z_l$ & Uniform over [0.0, $z_s$] \\
  Lens mass & $m_l$ & Mass-weighted Chabrier (2003) IMF \\
  Tranverse velocity & $v_t$ & 300 km s$^{-1}$  [0, 400] \\
  Minimum impact parameter & $u_0$ & Uniform over [0.063, 0.63] \\
  Time sampling & - & CRTS first-order time difference \\
  Second mass time offset & - & Uniform over [-500, 500] \\
\hline
\end{tabular}
\end{table*}

\begin{figure}
\centering
\caption{The skewness of the Weibull fit to the flare candidates against the duration of the flare.}
\label{skewdur}
\includegraphics[width = 3.5in]{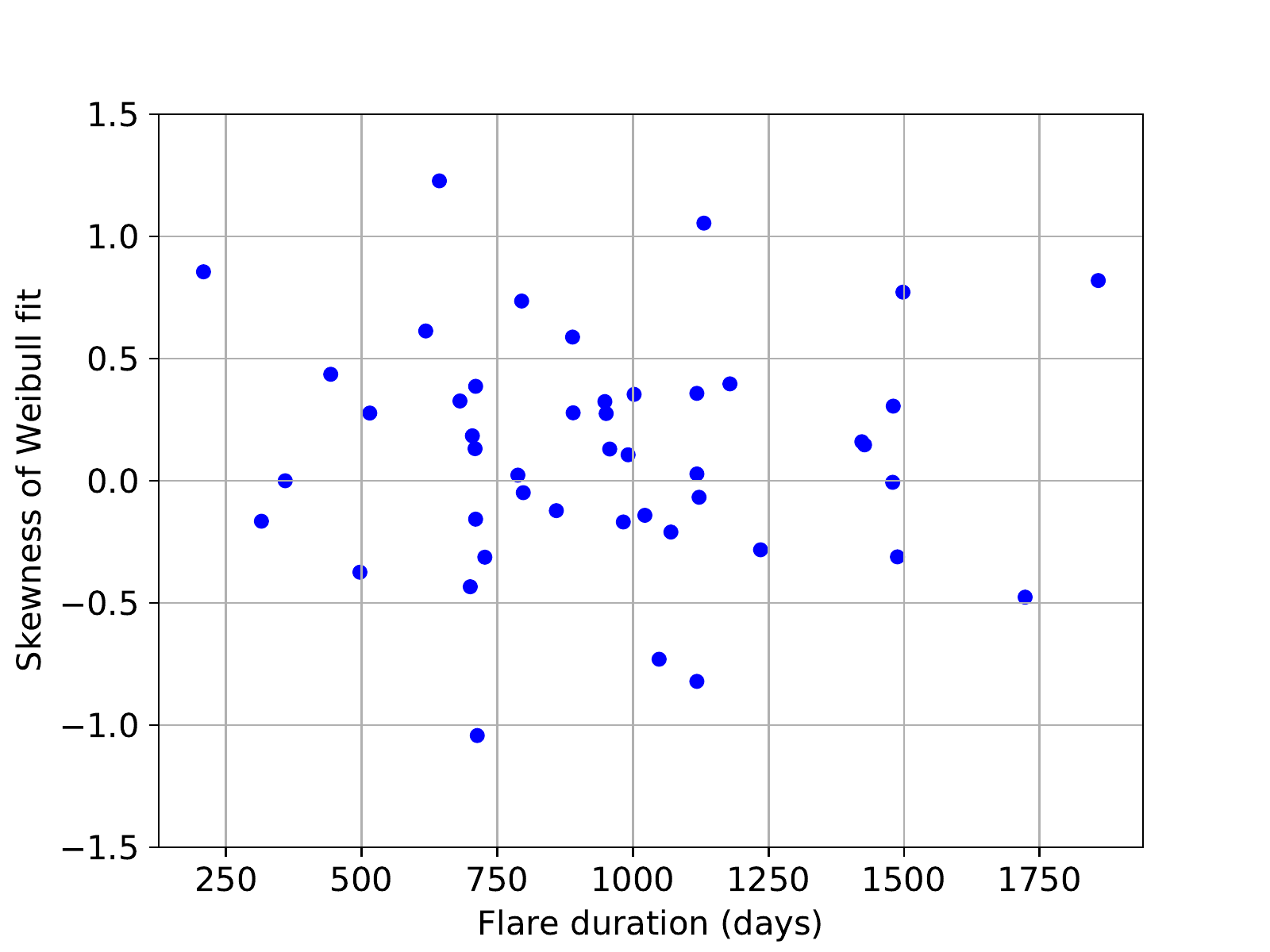}
\end{figure}

Collectively the single-point single lens model is not a good match to the flare candidates we have identified: our sources consistently have larger scale and smaller shape Weibull parameters, indicating more broader and less peaked flares. We note as well that L16 found that only three of their objects showed behavior that was consistent with a simple point-lens point source system. However, there are a number of the flare candidates whose Weibull parameterization overlaps with that of the single-point single lens model. We have therefore determined best-fit single-point single-lens models for each flare using MCMC (via the Python {\it emcee} package, \citealt{foreman13}) and the priors described above. For the lens redshift, we assume a uniform prior over the range $[0, z_s]$, where $z_s$ is the source redshift. When the source redshift is unknown, we assume a fiducial value of $z = 1.309$, which is the median redshift of the CRTS data set. Table~\ref{lenspars} gives the best-fit parameters for the eleven candidates which are well fit $(\chi^2_{red} < 1.8$) by a single-point single-lens model. 

In section~\ref{results}, we compared the DRW model parameter values for light curves with and without the identified flares. \cite{bruce16} performed a similar analysis on two of the L16 lensing candidates, finding that DRW model parameters were atypical for quasars for the observed data but more typical after subtraction of a microlensing model. We have also checked the effect on DRW model parameter values of subtracting the best-fit single-point single-lens models from the eleven lensed candidates. We find that the residual light curves of nine objects ($\chi^2_{red} < 1.55)$ are more consistent with a DRW variability. However, two objects (with $\chi^2_{red} > 1.55$) show little difference in their DRW parameter values between the original light curve and the lens model subtracted curve. This supports the microlensing hypothesis for the nine sources but we defer a fuller analysis of them and their spectra to a future paper.

\begin{table*}
  \centering
  \caption{The best-fit MCMC parameters for single-point single-lens models }
  \label{lenspars}
  \begin{tabular}{llllllll}
  \hline
ID & $z$ & $m_l$& $v_t$ & $t_0$ & $z_l$ & $u_0$ & $\chi^2_{red}$ \\
 & & $(M_\odot)$ & (km s$^{-1}$) & (day) & & \\
  \hline
J093941$+$100706 & 0.328 & 0.69 & 540 & 701 &  0.16 & 0.80 & 0.771 \\
J150448$-$250702 & -  &  2.09 & 380 & 582 & 0.61 & 0.47 & 0.894 \\
J213007$-$015556 & 0.290 &  0.66 & 130 & 1125 &   0.13 & 0.70 & 1.272 \\
J144321$+$344940 & 0.749 &  1.08 & 300 & 557 &   0.32 & 0.66 & 1.308 \\
J110033$+$160808 & (1.090) & 0.79 & 210 & 1205 &  0.44 & 0.59 & 1.311 \\
J030328$-$033821 & 0.703 & 12.2 & 609 & 778 &   0.415 & 0.522 & 1.324 \\
J004133$+$212841 & 0.343 & 1.33 & 480 & 1066 &   0.17 & 0.84 & 1.328 \\
J113412$+$192226 & 0.843 &  7.12 & 597 & 476 &   0.365 &  0.414 & 1.442 \\
J232638$+$000524 & 1.031 & 1.38 & 470 &  475 &  0.43 & 0.52 & 1.545 \\
J010234$+$050853 & 1.432 & 8.59 &  315 & 542 & 1.073 & 0.554 & 1.595 \\
J150032$+$044247 & 0.971 &  5.67 & 295 & 412 &   0.480 & 0.480 & 1.672 \\
\hline
\end{tabular}
\end{table*}

L16 identified CRTS transient data for sixteen of their sources (four of which are AGN); however, CRTS archival data is available for 58 of their AGN, including fourteen of the fifteen identified in Stripe 82. Of these we have identified nine that meet our criteria for a major AGN flare (three of these are also included in \cite{bruce16} as lensed sources). It should be noted that L16 actually distinguish between four categories of AGN light curve shape: rising, falling, peaked, and complex. The rising and falling types are more consistent with behavior associated with changing-look quasars rather an outburst event and we defer further discussion of this to a subsequent paper. 
The Weibull characterizations for the nine events (see Table~\ref{lawrence}) have only partial overlap with the distribution from simulated lens flares, suggesting again that lensing is the physical explanation for the flare in only some of the sources and that a number of different physical mechanisms are contributing overall to this phenomena.

\begin{table*}
  \centering
  \caption{The list of AGN flare candidates from Lawrence et al (2016) meeting our criteria.}
  \label{lawrence}
  \begin{tabular}{lllllllll}
  \hline
  ID & CRTS ID & $V_{med}$ & $z$ & $\Delta T$ & $amp_{max}$  & $a$ & $s$ & Total energy \\
       & & & & (day) & (mag) & & &  (erg) \\
  \hline
 J025633$+$370712 & 1138013014735 &  19.70 & 0.000 & 635 & 1.49 & 2.52 & 364 &  $4.20 \times 10^{37}$ \\
 J083714$+$260932 & 1126042018137  &  20.07 & 0.000 & 683 & 1.65 & 1.03 & 311 &  $1.89 \times 10^{40}$ \\
 J090514$+$503628 & 1149033050415  &  19.91 & 1.290 & 438 & 0.70 & 1.08 & 6110 &  $3.74 \times 10^{51}$ \\
 J094511$+$174544 & 2117130011595 &  20.58 & 0.758 & 1172 & 0.85 & 1190 & 480000 &  $1.84 \times 10^{51}$ \\
 J103837$+$021119 & 1101057044911  &  19.62 & 0.620 & 735 & 1.06 & 7.71 & 2530 & $1.24 \times 10^{51}$ \\
 J104617$+$553336 & 1155035041492 &  20.22 & 0.000 & 470 & 1.70 & 2.87 & 268 &  $8.52 \times 10^{41}$ \\
 J105501$+$330002 & 1132051039346 &  19.29 & 0.417 & 833 & 1.04 & 3.96 & 722 &  $7.46 \times 10^{44}$\\
 J142232$+$014026 & 1101077043951 &  19.45 & 1.079 & 735 & 0.67 & 1.59 & 1670 &   $2.21 \times 10^{51}$ \\
 J150210$+$230915 & 1123074014056 &  19.95 & 0.630 & 544 & 1.64 & 3.19 & 1690 &  $6.67 \times 10^{48}$ \\
 \hline
\end{tabular}
\end{table*}

\subsection{Superluminous supernovae}

During the past decade dozens of superluminous supernova have been discovered by wide-field transient surveys such as CRTS, PTF and PS1.  Supernovae are defined as superluminous (SLSN) when they reach 
$M_V < -21$ \citep{drake10, galyam12}. The most luminous SLSNe observed have $M_V \sim -22.5$ \citep{galyam12}. In contrast,  regular supernovae typically peak in the range $-17 < M_V < -20$ \citep{richardson02}.

SLSNe are generally divided into the hydrogen poor (SLSN-I) and hydrogen rich (SLSN-II) types \citep{galyam12}.  The origin of SLSN-I  events is not yet certain but in at least some cases are likely to be due to Wolf-Rayet stars \citep{tadia16}. In contrast, SLSN-II have been attributed to more luminous versions of type-IIn SN, which themselves are believed to be due to the end stages of luminous blue variables (LBVs) with massive circumstellar matter (CSM) envelopes. Both types of SLSNe have been measured to emit more than $10^{51}$ ergs of energy \citep{smith07, drake10, quimby11, rest11} and are thus within the range of almost all the flares shown here. Furthermore, examples of SLSN have been discovered up to redshifts of $z=3.9$ \citep{cooke12}. So the distances to these flaring sources are not exceptional.

One of the clearest signatures of a supernova is a smooth rising light curve followed by a typically much longer decline. The timescale of the rise varies between supernova types. For example, type-Ia supernovae have rise times of only two weeks, whereas type-IIn can take months. Nevertheless, the overall shape of the rising curves is driven by expansion and has long been known to be similar for differing types of supernovae 
\citep{wheller90}. Our fits to many of the flares presented appear consistent with the asymmetric
shapes of supernova light curves. However, the average timescale of the flares is $\sim 900$ days. This is inconsistent with those observed for either regular supernovae or SLSN-I (which generally last $\la$200 
rest-frame days). On the other hand, type-IIn supernovae and the related SLSN-II can last for years \citep{mahabal09, fox15}. Nevertheless, although the total energies of the flares are within the range 
of SLSNe-II, two thirds of the flares have peak absolute magnitudes brighter than $M_V = -23$. The combination of long timescales and high peak magnitudes suggest that if the flares are due to SLSNe-II, they would have to be an extreme tail.

For SLSNe-II the timescales of the event depends on both the extent and configuration of the circumstellar material (CSM) environment in which they reside \citep{chatzopoulos13}. It is possible that star formation within the AGN disk could lead to the production of massive stars \citep{levin07}. The short life times of such stars in turn is expected to produce type-II supernovae. If the ejecta from these events could interact with both the CSM from their own outbursts as well as the gas from the AGN disk, the events may be more luminous and longer lived than historical SN that have overwhelmingly been observed away from the cores of galaxies in order to avoid AGN.

One possible example of a SLSN-II associated with an AGN is CSS100217:102913+404220 \citep{drake11}. However, the presence of an AGN, combined with the similarity of AGN and type-IIn spectra, has meant that the event could not be firmly identified as a SLSN. Overall, it seems very unlikely that all the flares we observe could be due to SLSNe-II. For example, many of the flares have longer rise times than declines. Such events are yet to be observed among known supernovae. The recent discovery that the putative most luminous SLSN-I (ASASSN-15lh, \cite{dong16}) is more likely a tidal disruption event (TDE) than a supernova \citep{brown16,leloudas16} also suggests that very bright flares need not be due to SLSNe.

\subsection{Stellar Mass Binary Black Hole Merger}

An intriguing potential cause of the observed AGN flaring is a stellar mass binary black hole merger within the dense medium of an AGN accretion disk.  The Laser Interferometer Gravitional-Wave Observatory (LIGO) has recently reported the exciting detection of gravitational waves from multiple double stellar-mass black hole binary systems, the first due to the merger of a $36^{+5}_{-4}\, M_\odot$ and a $29 \pm 4 \, M_\odot$ black hole (GW150914; \cite{abbott16a}), and the second due to the merger of a $14.2^{+8.3}_{-3.7}\, M_\odot$ and a $7.5 \pm 2.3 \, M_\odot$ black hole (GW151226; \cite{abbott16b}).  Stellar-mass black holes are expected to sink towards the nuclei of galaxies due to dynamical friction with stars.  Some of these stellar-mass black holes will have formed in binaries, while others will form due to dynamical interactions in dense stellar systems, such as found in galactic nuclei.  For galaxies hosting active galactic nuclei, many of these black hole binary systems
will migrate into the associated accretion disk \citep{mckernan17}.

\cite{bartos16} investigate the time scales for both the orbital alignment of stellar mass black hole orbits with the accretion disk, as well as the accelerated merger time scale of stellar mass black hole binaries within an accretion disk.  For their fiducial model of a 75 $M_\odot$ black hole (which is expected to be essentially
equivalent to black hole binary system totaling that mass) and a $10^6\, M_\odot$ central supermassive black hole, they find that a significant fraction ($\sim 12\%$) of stellar mass black holes / black hole binary systems will align themselves with the accretion disk within $10^7$~yr.  This fraction rises to $\sim 43\%$ within
$10^8$~yr.  In the dense medium of an accretion disk, the binaries will then merge at an accelerated pace as compared to isolated stellar mass black hole binaries, with mergers expected within $\sim 10^6$~yr, first primarily driven by dynamical friction (at early stages) and later due to gravitational radiation (in the final stages).  

While in the accretion disk, the black holes are also expected to accrete gas from the disk at levels well above the Eddington rate, producing significant high-energy emission.  While \cite{bartos16} primarily investigates the gravitational wave and high-energy electromagnetic signatures of such events, we make the speculative suggestion that these optical AGN flares could be due to secondary emission related to stellar mass black hole binaries in the dense environment of an accretion disk, either during their pre-merger, super-Eddington accretion stages, or, perhaps they are related to the black hole merger event, though we note that the latter seems less likely given the discrepancy in the timescales.

It is also possible that the flaring might be related to a single stellar-mass black hole migrating through an AGN accretion disk due to torques from gas that is co-rotating and at (inner and outer) Lindblad resonances \citep{mckernan14}. Although the migration is a long term ($\sim 10^6$ years) process, a sufficiently massive migrator (compared to the co-moving disk gas) could open a gap in the gas disk that would act as a migration trap. A stalled migrator at such a trap may generate a tidal bulge in the gas exterior to its orbit, if the disk is relatively viscous and gas continues to flow inward. If the bulge suddenly collapses, it might generate a flare. A plunging low angular momentum retrograde orbiter would also generate a flare as it ploughs through the inner disk by dropping low angular momentum gas onto the central black hole.

\subsection{Slow TDEs}

A typical tidal disruption event shows a fast rise to a peak luminosity of $\sim 10^{44} $ergs followed by a decay following $t^{-5/3}$ with a timescale to consume half the material of $t_{1/2} \sim 120$ days. Such events are also more frequent around lower mass black holes $(\sim 10^{6-7} M_\odot)$ because of steeper force gradients. 
Although these characteristics are different from the events described here, \cite{guillochon15} have shown that
relativistic precession arising from black hole spin can prevent the debris stream from the TDE from self-intersecting until after many windings. This introduces a delay, possibly of several years, before the flare becomes observable and a shallower power-law decline closer to $t^{-1}$ for lower mass black holes. Such TDEs will be sub-Eddington at the peak and so will have been missed in current searches. 

We have determined the best-fitting decay profile for each flare via a Thiel-Sen fit in log-space to the flare flux (see Table~\ref{candidates}). We find eight candidates with flares characterized by a decaying exponent between -1 and -1.7, which represents the expected range. One of these (J213007-015556) is also a lensing candidate which we exclude as a TDE candidate as the flare profile is too symmetrical. Only two of the remaining seven sources (J005448+225123 and J010234+050853) have sufficiently short rise times to be considered a viable TDE event(the other five candidates all have a longer rise time than fall time which is not the expected profile). From the spectra of these quasars, we measure Mg II equivalent widths of 1100 km s$^{-1}$ (J005448+225123) and 4500 km s$^{-1}$ (J010234+050853) respectively, which give virial black hole masses of $\log_{10}(M / M_{\odot}) = 7.8$ and 8.9 using \cite{kozlowski16b}. Assuming that the viscous time for the accretion disk is 100 times longer than the orbital period, \cite{guillochon15} find that the majority of events associated with black holes below a fiducial mass of $\log_{10}(M / M_{\odot}) = 7.0$ are slowed. A longer viscous time leads to a higher fiducial mass and more slow TDEs around higher mass black holes. The less massive of our two candidates, J005448+225134, has the shallower decay slope and longer event duration but further modeling is required to see if slow TDEs are a viable explanation for some of these events.

\section{Conclusions}

We have identified 51 quasars which over the past decade have exhibited a major flaring event that is statistically distinct from their expected stochastic variability. The event typically lasts about 900 days (in the observed frame) and has a median peak amplitude of $\Delta m = 1.25$ mag. The flares have one of three distinct profiles: symmetric, fast rise exponential decay, and slow rise, fast decay. \cite{lawrence16} has proposed that many such events are attributable to microlensing. A single-point single lens model provides a good description for nine of the flares but we suggest that the rest are due to some form of explosive stellar activity in the accretion disk: either a superluminous supernova, a (slowed) tidal disruption event, or even a merger event. Further modeling, both of more complex lensing geometries and stellar-initiated activity within the accretion disk of an AGN, will help to understand  these events.

Followup observations, both spectroscopic and multiwavelength, would also help to discriminate between different models. The long baseline of these events means that there may be serendipitous observations in existing archives and we will consider this in a subsequent paper. The increasing number of sky surveys and sky coverage per night also means that more of these events should be discovered in future. We estimate the rate of a $\Delta m =1$ magnitude event with a lifetime of $\sim 1000$ days in the AGN population to be $\sim 10^{-5}$ yr$^{-1}$ sr$^{-1}$. 
A statistically useful sample should therefore be feasible within the first few years of LSST.

Although these events may offer more insight into the structure of the accretion disk, e.g., constraints on the size of particular regions from microlensing, they seem to be a distinct class of phenomenon from the more general 
variability seen in AGN. Theye are true outliers rather than representing the tail of any distribution. One possibility, however, is that these types of explosive events seed more general AGN variability by creating matter streams, shock fronts, and inhomogeneities in the (inner) accretion disk. This would then support the type of variability mechanisms proposed by \cite{aretxaga97} and \cite{torricelli00}. Again further modeling is needed to see whether the expected observational signatures match what is actually seen.

\section*{Acknowledgements}

We thanks Hans Meusinger for providing the data for Sharov 21 in electronic form, Chelsea MacLeod for the predicted cumulative magnitude distributions, Alastair Bruce for clarifying the L16 lens fits, and Barry McKernan for discussions on accretion disk physics. We also thank the anonymous referee for useful comments. 

This work was supported in part by the NSF grants AST-1413600 and AST-1518308. The work of DS was carried out at Jet Propulsion Laboratory, California Institute of Technology, under a contract with NASA.

This work made use of the Million Quasars Catalogue.

Funding for SDSS-III has been provided by the Alfred P. Sloan Foundation, the Participating Institutions, the National Science Foundation, and the U.S. Department of Energy Office of Science. The SDSS-III web site is http://www.sdss3.org/.

SDSS-III is managed by the Astrophysical Research Consortium for the Participating Institutions of the SDSS-III Collaboration including the University of Arizona, the Brazilian Participation Group, Brookhaven National Laboratory, Carnegie Mellon University, University of Florida, the French Participation Group, the German Participation Group, Harvard University, the Instituto de Astrofisica de Canarias, the Michigan State/Notre Dame/JINA Participation Group, Johns Hopkins University, Lawrence Berkeley National Laboratory, Max Planck Institute for Astrophysics, Max Planck Institute for Extraterrestrial Physics, New Mexico State University, New York University, Ohio State University, Pennsylvania State University, University of Portsmouth, Princeton University, the Spanish Participation Group, University of Tokyo, University of Utah, Vanderbilt University, University of Virginia, University of Washington, and Yale University.






\begin{landscape}
\begin{table}
  \centering
  \caption{The list of AGN flare candidates. Sources for which we have obtained a spectroscopic redshift are marked with a ``*''; photometric redshifts are quoted in parentheses. $\Delta T$ is the observed timespan of the light curve. $a$ and $s$ are the fitted Weibull parameters. Decay is the best fit exponent to the flare profile. Note that sources without a redshift do not have a redshift-corrected total energy value.}
  \label{candidates}
  \begin{tabular}{lllllllllllll}
  \hline
  ID & CRTS ID & RA & Dec & $V_{\mathrm{med}}$ & $z$ & $\Delta T$ & $amp_{\mathrm{max}}$  & $a$ & $s$ & Peak. abs. & Total energy & Decay \\
       & & & & & & (day) & (mag) & & & mag. &  (erg) & \\
  \hline
 J000727$-$132644 & 1012001011298 & 00 07 27.65 & -13 26 44.16 & 18.86 & 0.699* & 2891 & 1.03 & 3.07 & 1610 & -23.5 & $3.04 \times 10^{51}$ & -1.3 \\
  J002237$+$000519 & 2100006002614 & 00 22 37.91 & +00 05 19.14 & 20.47 & 1.373 & 2996 & 1.15 & 2.42 & 
1510 &  -23.5  & $3.09 \times 10^{51}$ & -1.2 \\
 J002748$-$055559 & 1007003045309 & 00 27 48.39 & -05 55 59.41 & 18.60 & 0.429* & 2886 & 0.90 &2.66 & 547 &  -25.0 & $4.52 \times 10^{50}$ & -6.0 \\
 J004133$+$212841 & 1121004039565 & 00 41 33.26 & +21 28 41.52 & 18.37 & 0.343 & 3031 & 0.89 & 5.45 & 929 & -22.0 & $5.90 \times 10^{50}$ & -3.7 \\
 J005448$+$225123 & 1123005006362 & 00 54 48.53 & +22 51 23.76 & 17.76 & 0.744* & 1954 & 1.02 & 2.69 & 529 &  -25.3 & $1.22 \times 10^{52}$ & -1.0 \\
 J010032$+$042408 & 2104014001615 & 01 00 32.04 & +04 24 08.42 & 19.31 & 0.721 & 3024 & 0.88 & 12.7 & 3690 & -23.5  & $3.10 \times 10^{51}$ & -1.6 \\
 J010234$+$050853 & 2104015017473 & 01 02 34.44 & +05 08 53.41 & 20.14 & 1.432* & 3024 & 1.15 &  2.5 & 584 &  -23.7 & $2.82 \times 10^{51}$ & -1.5 \\
 J012145$+$045504 & 1104008037628 & 01 21 45.50 & +04 55 04.80 & 18.44 & 0.840 & 2881 & 2.82 & 1.93 & 1180 &  -25.5 & $2.90 \times 10^{52}$ & -2.5 \\
 J012612$+$113016 & 2111020015465 & 01 26 12.34 & +11 30 16.20 & 19.82 & 0.800 & 2942 & 1.18 & 7.17 & 4090 & -23.5  & $3.56 \times 10^{51}$ & -1.8 \\
 J022014$-$072859 & 1007013018204 & 02 20 14.57 & -07 28 59.34 & 17.03 & 0.213 & 2922 & 0.78 & 59.6 & 12000 & -22.0 & $1.14 \times 10^{51}$ & -0.3 \\
 J023439$+$010742 & 1101014022249 & 02 34 39.07 & +01 07 42.67 & 19.50 & 0.277 & 2881 & 1.96 & 3.82 & 286 & -21.5 & $5.22 \times 10^{50}$ & -\\
 J025411$+$255324 & 1126015012060 & 02 54 11.02 & +25 53 24.72 & 18.84 & 0.331 & 2877 & 1.57 &2.07 & 718 &  -22.3 & $8.13 \times 10^{50}$ & -1.9\\
 J030328$-$033821 & 1004017038032 & 03 03 28.63 & -03 38 21.59 & 19.09 & 0.703* & 2963 & 1.30 &  $>10^5$ & $>10^5$ & -24.0 & $2.10 \times 10^{51}$ & -3.9 \\
 J030606$+$192643 & 2118041030282 & 03 06 06.67 & +19 26 43.08 & 20.55 & 0.522* & 2939 & 1.85 &5.44 & 2270 & -22.5 & $1.66 \times 10^{51}$ & -2.5 \\
 J081333$+$183446 & 2118109005517 & 08 13 33.60 & +18 34 46.20 & 20.42 & 0.897* & 2920 & 1.34 & 4.37 & 591 & -24.5 & $1.42 \times 10^{51}$ & -5.3\\
 J083027$+$203652 & 2121111001208 & 08 30 27.12 & +20 36 52.20 & 19.73 & 1.310 & 2936 & 0.84 &3.51 & 954 &  -& $4.95 \times 10^{51}$ &-1.7 \\
 J084339$-$015109 & 1001047028669 & 08 43 39.60 & -01 51 09.22 & 17.88 & 0.809 & 3124 & 1.07 & 4.44 & 986 & -24.5 & $5.63 \times 10^{51}$ &-2.6 \\
 J090347$+$151818 & 2115122003985 & 09 03 47.76 & +15 18 18.72 & 20.43 & 1.413* & 2968 & 1.44 &  $>10^5$ & $>10^5$ & -23.0 & $4.15 \times 10^{51}$ &-2.6 \\
 J090612$+$272347 & 1126045043734 & 09 06 12.24 & +27 23 47.40 & 18.74 & 0.920* & 2211 & 1.66 & 2.58 & 1310 &  -21.5 & $6.46 \times 10^{51}$ & -1.1\\
 J092407$+$615626 & 1160026056310 & 09 24 07.68 & +61 56 26.52 & 18.10 & 0.205 & 2653 & 0.83 &4.69 & 661 & -& $3.34 \times 10^{50}$ & -0.2 \\
 J092415$+$164902 & 2116126012459 & 09 24 15.36 & +16 49 02.28 & 18.93 & 0.352 & 2927 & 0.00 & 1.85 & 421 &  - & $5.35 \times 10^{50}$ & -2.8 \\
 J093941$+$100706 & 1109052036713 & 09 39 41.04 & +10 07 06.60 & 18.80 & 0.328 & 2867 & 0.75 & 6.64 & 407 & -21.7  & $2.51 \times 10^{50}$ & -0.4\\
 J094608$+$351222 & 1135044028613 & 09 46 08.40 & +35 12 22.68 & 17.19 & 0.119 & 2954 & 0.90 &2.03 & 286 &  -20.8 & $3.14 \times 10^{50}$ &-2.8 \\
 J094806$+$031801 & 1104053011095 & 09 48 06.48 & +03 18 01.44 & 17.83 & 0.207 & 2961 & 1.11 &  2.44 & 361 &  -22.0  & $8.82 \times 10^{50}$ &-3.7 \\
 J094932$+$241553 & 1123049035019 & 09 49 32.64 & +24 15 53.28 & 18.79 & 1.123 & 2977 & 1.10 & 3.12 & 361 &  -25.0 & $8.09 \times 10^{51}$ & - \\
 J101524$+$145840 & 1115054017365 & 10 15 24.72 & +14 58 40.80 & 17.97 & 1.102 & 3122 & 2.09 &  1.71 & 996 &  -26.0  & $2.60 \times 10^{52}$ & -2.3 \\
 J102515$+$003640 & 1101056013653 & 10 25 15.36 & +00 36 40.79 & 19.37 & 0.817 & 2492 & 1.36 & 2.95 & 558 &  -24.0 & $2.90 \times 10^{51}$ & -2.2 \\
 J102912$+$404220 & 1140044024955 & 10 29 12.48 & +40 42 20.16 & 17.50 & 0.147 & 2739 & 1.73 & 1.52 & 189 &  -22.5 & $8.25 \times 10^{50}$ & -4.0 \\
 J103146$+$072411 & 2107146024798 & 10 31 46.80 & +07 24 11.30 & 20.16 & (1.064) & 2936 & 1.24  & 1.75 & 7450 &  -23.5 & $4.40 \times 10^{51}$& -0.8 \\
 J105230$+$182043 & 1118056030240 & 10 52 30.48 & +18 20 43.08 & 19.46 & 0.693 & 3127 & 2.45 & 2.7 & 816 &  -24.5 & $6.58 \times 10^{51}$ & -4.0 \\
 J110033$+$160808 & 1115057039761 & 11 00 33.84 & +16 08 08.16 & 18.76 & (1.090) & 3127 & 1.99 &5.21 & 1310 & -25.0  & $8.72 \times 10^{51}$ &- 2.3\\
 J111306$-$011845 & 1001060034907 & 11 13 06.96 & -01 18 45.07 & 18.89 & 0.981 & 2987 & 1.53 & 16.8 & 7150 & -24.8 & $8.98 \times 10^{51}$ & -2.4 \\
 J113008$+$005054 & 2100160020745 & 11 30 08.88 & +00 50 55.00 & 19.71 & 2.100 & 2924 & 1.60 & 1.81 & 1650 &  -25.8 & $2.12 \times 10^{52}$ & -1.9 \\
 J113412$+$192226 & 1118060051368 & 11 34 12.48 & +19 22 26.76 & 19.73 & 0.843 & 2576 & 1.76 &  2.57 & 535 &  -24.0 & $1.83 \times 10^{51}$ & -2.7 \\
 J120715$-$023329 & 2002169018019 & 12 07 15.36 & -02 33 29.30 & 20.51 & (1.210) & 2833 & 2.46 &  3.63 & 1660 & -24.3 & $8.09 \times 10^{51}$ & -2.8 \\
J123613$+$001733 & 1101068005915 & 12 36 13.68 & +00 17 33.79 & 19.55 & 0.590 & 2979 & 1.86 & $>10^5$ & $>10^5$ & -23.0 & $1.83 \times 10^{51}$ & -3.1 \\
 J124730$-$014227 & 1001069023492 & 12 47 30.96 & -01 42 27.22 & 18.28 & 0.347 & 2984 & 0.67 & 4.18 & 1360 & -22.3 & $7.92 \times 10^{50}$ & -1.0 \\
 J131150$+$192053 & 1118068053018 & 13 11 50.64 & +19 20 53.16 & 17.66 & 0.398 & 2994 & 0.56 &  3.12 & 521.0 &  -25.0 & $8.86 \times 10^{50}$ & -1.8\\
 J140710$-$122309 & 2012192015243 & 14 07 10.32 & -12 23 09.24 & 20.18 & 0.659 & 2914 & 1.23 &  3.49 & 1140 &  -22.7  & $1.30 \times 10^{51}$ & -1.8 \\
 J141828$+$354248 & 1135064040384 & 14 18 28.56 & +35 42 48.96 & 19.66 & 2.100 & 2984 & 2.53 & 4.42 & 1240 & -25.0 & $1.08 \times 10^{53}$ & -10.2 \\
 J144321$+$344940 & 1135066022284 & 14 43 21.12 & +34 49 40.44 & 18.48 & 0.749 & 2980 & 1.03 &  3.2 & 631 &  -24.0 & $3.26 \times 10^{51}$ & -2.1 \\
 J145116$+$343542 & 1135066016768 & 14 51 16.08 & +34 35 42.36 & 18.90 & (1.475) & 2935 & 1.29 & 2.62 & 1750 &  -25.0 & $1.12 \times 10^{52}$ & -1.8 \\
 J150032$+$044247 & 1104081049591 & 15 00 32.88 & +04 42 47.20 & 19.64 & (0.971) & 2990 & 1.42 & 697 & 651 & -24.0  & $1.34 \times 10^{51}$ & - \\
 J150448$-$250702 & 3025103031853 & 15 04 48.96 & -25 07 03.00 & 19.06 & 0.000 & 2868 & 1.83 & 6.01 & 1050 & -25.8 & $1.73 \times 10^{37}$ & -3.3 \\
 J152205$+$102125 & 1109082051016 & 15 22 05.04 & +10 21 25.20 & 19.48 & (0.903) & 2987 & 2.51 & 2.7 & 477 &  - & $1.14 \times 10^{52}$ & -1.9 \\
 \hline
   \end{tabular}
 \end{table}
\end{landscape}

\begin{landscape}
\begin{table}
  \centering
  \contcaption{}
\begin{tabular}{lllllllllllll}
  \hline
  Id & CRTS ID & RA & Dec & $V_{med}$ & $z$ & $\Delta T$ & $amp_{max}$  & $a$ & $s$ & Peak abs. & Total energy  & Decay\\
       & & & & & & (day) & (mag) & & & mag. &  (erg) & \\
  \hline 
 J161542$+$024651 & 1101087096238 & 16 15 42.72 & +02 46 51.13 & 18.04 & 0.326* & 3071 & 1.28 &2.51 & 582 &  -23.0 & $2.90 \times 10^{51}$ & -1.8 \\
 J213007$-$015556 & 1001115026824 & 21 30 07.92 & -01 55 56.93 & 18.23 & 0.290 & 3111 & 0.86 & 4.7 & 1190 & -22.0 & $1.01 \times 10^{51}$ & -1.7 \\
 J223139$+$122107 & 1112119025810 & 22 31 39.84 & +12 21 07.92 & 19.27 & 0.603* & 3055 & 1.81 &2.34 & 296 &  -23.3 & $4.74 \times 10^{44}$ &-3.6 \\
 J224720$-$060525 & 2005315009619 & 22 47 20.88 & -06 05 25.87 & 19.26 & 1.669* & 3013 & 1.08 &  4.29 & 1310 & -26.0  & $2.36 \times 10^{52}$ & -3.6 \\
 J224736$-$082541 & 2008314006156 & 22 47 36.96 & -08 25 41.02 & 20.30 & 1.638* & 2986 & 1.21 &3.03 & 1420 &  -24.3 & $6.63 \times 10^{51}$ & -1.3 \\
 J232638$+$000524 & 2000326023025 & 23 26 38.16 & +00 05 24.65 & 20.32 & 1.031 & 2949 & 1.07 &  3.9 & 769 & -23.3 & $1.91 \times 10^{51}$ & -3.0 \\
 \hline
 Sharov 21 & - & 00 44 57.94 & +41 23 43.90 &  19.2 & 2.109 & 0 & 3.03 & 2190 & $>10^5$ & - & $1.39 \times 10^{52}$ & - \\ 
\hline \hline
  \end{tabular}
 \end{table}
\end{landscape}

\begin{figure*}
\centering
\caption{Light curves for flaring candidates. CRTS data (DR2: black, post-DR2: cyan) is shown; complementary data from the LINEAR (blue) \citep{sesar11} and PTF (red) \citep{rau09} surveys are included where available. The light curve for Sharov 21 is also shown for comparison. The line in each plot shows the best fit Weibull distribution to the identified flare. Note that this is relative to a linear model for the median activity of the source. Sources with a corresponding lens model in Fig. 17 are denoted by an asterisk in the source name.}
\label{lightcurves}
\includegraphics[width = 7.0in] {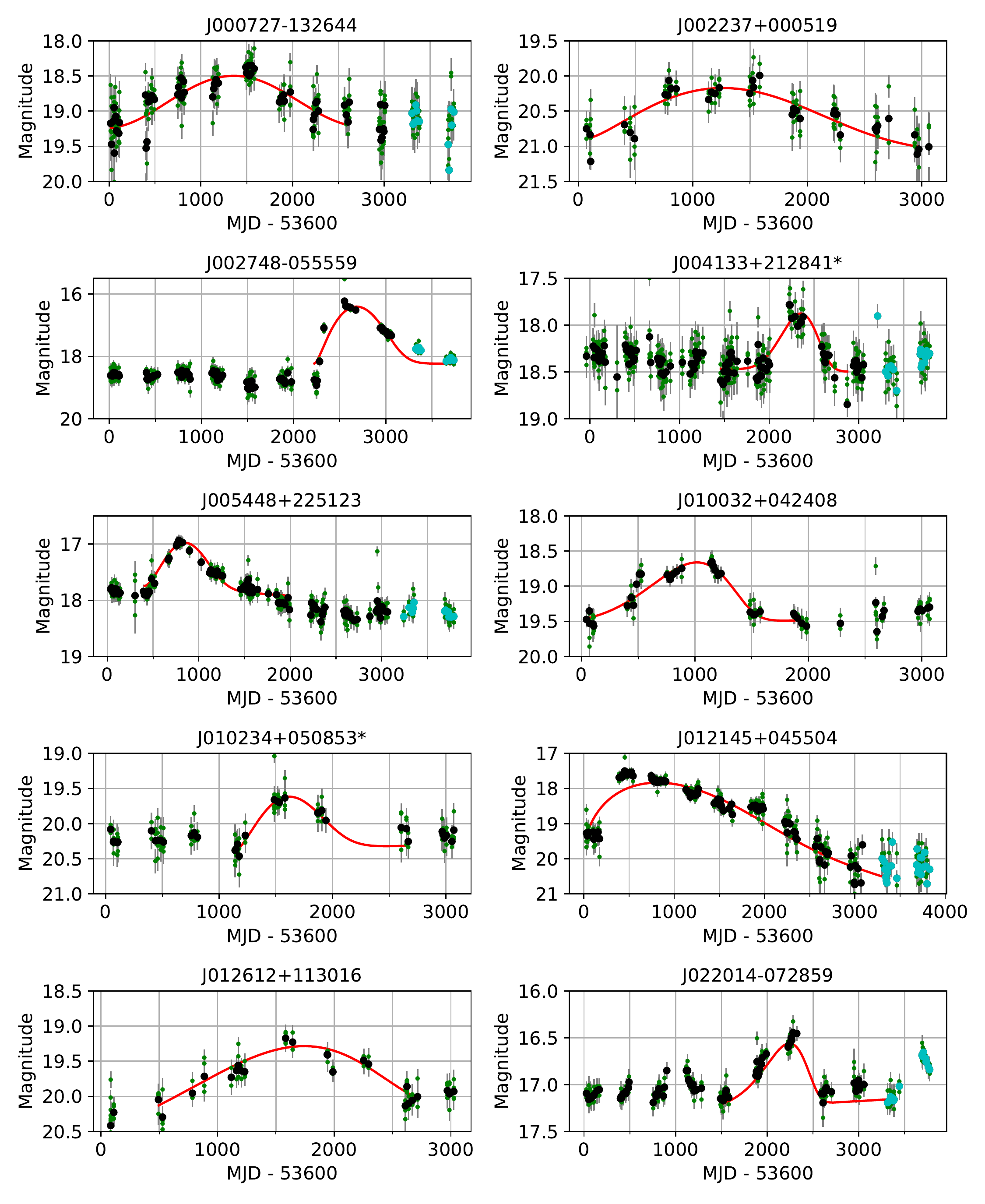}
\end{figure*}

\begin{figure*}
\centering
\contcaption{}
\includegraphics[width = 7.0in] {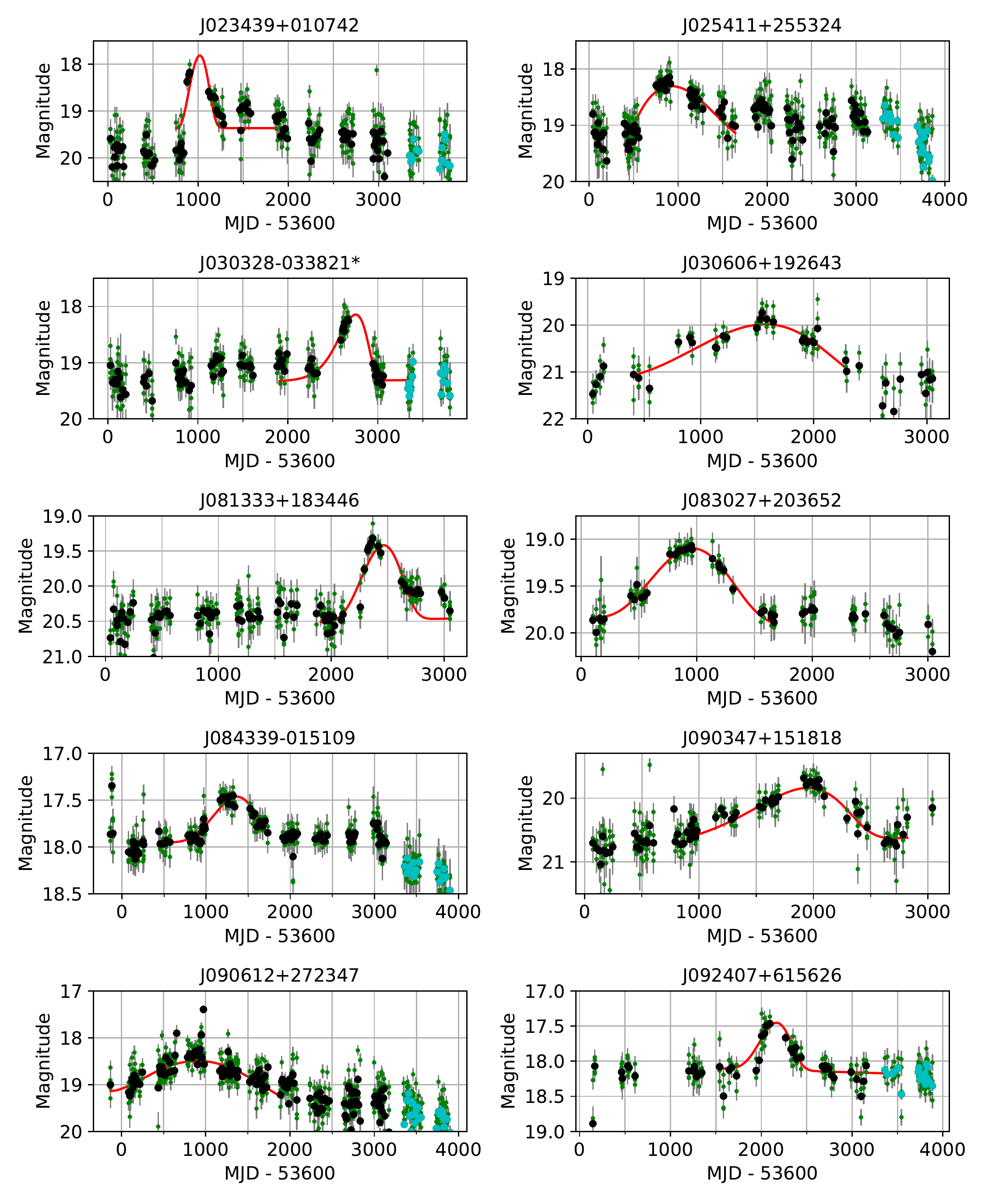}
\end{figure*}

\begin{figure*}
\centering
\contcaption{}
\includegraphics[width = 7.0in] {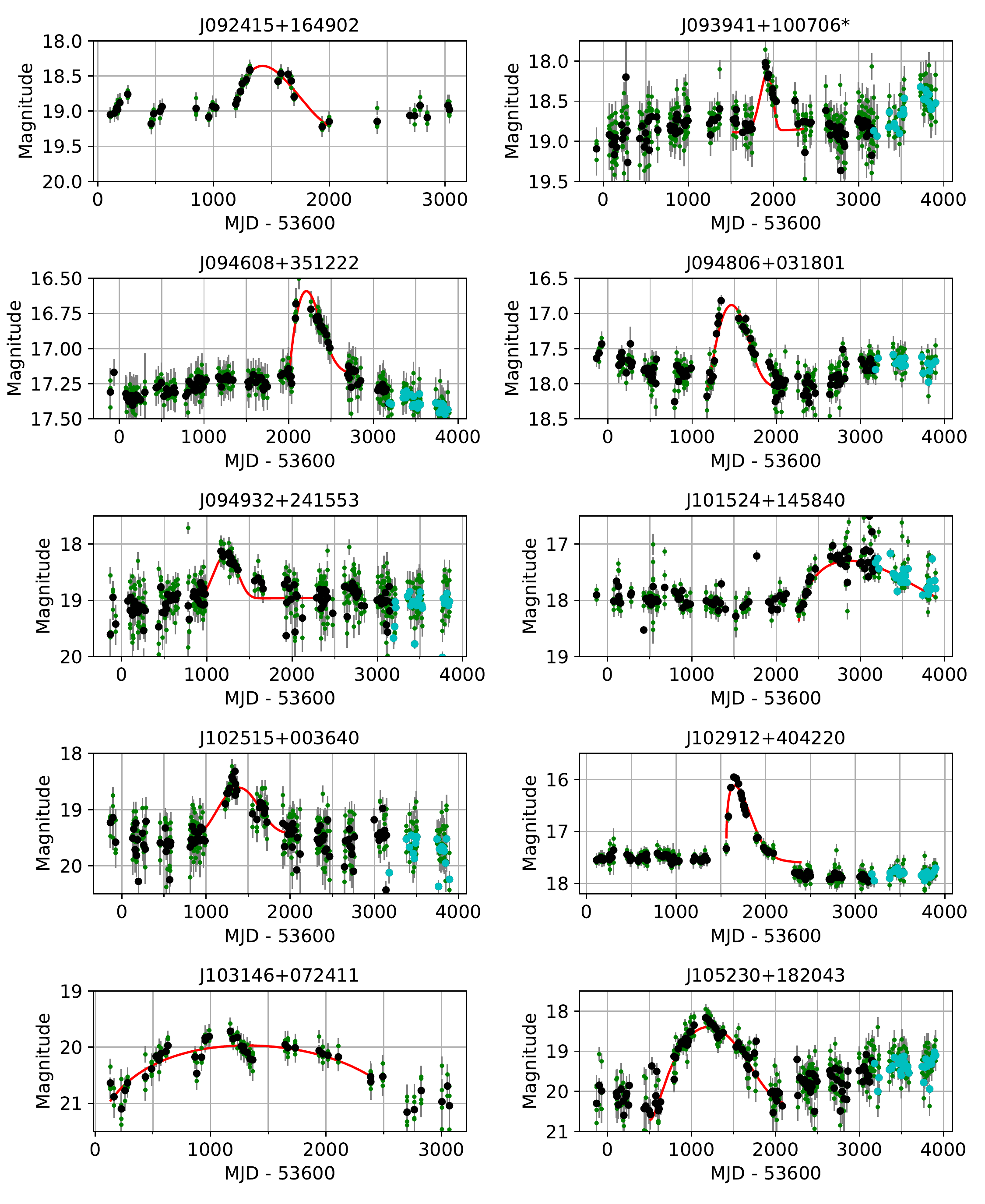}
\end{figure*}

\begin{figure*}
\centering
\contcaption{}
\includegraphics[width = 7.0in] {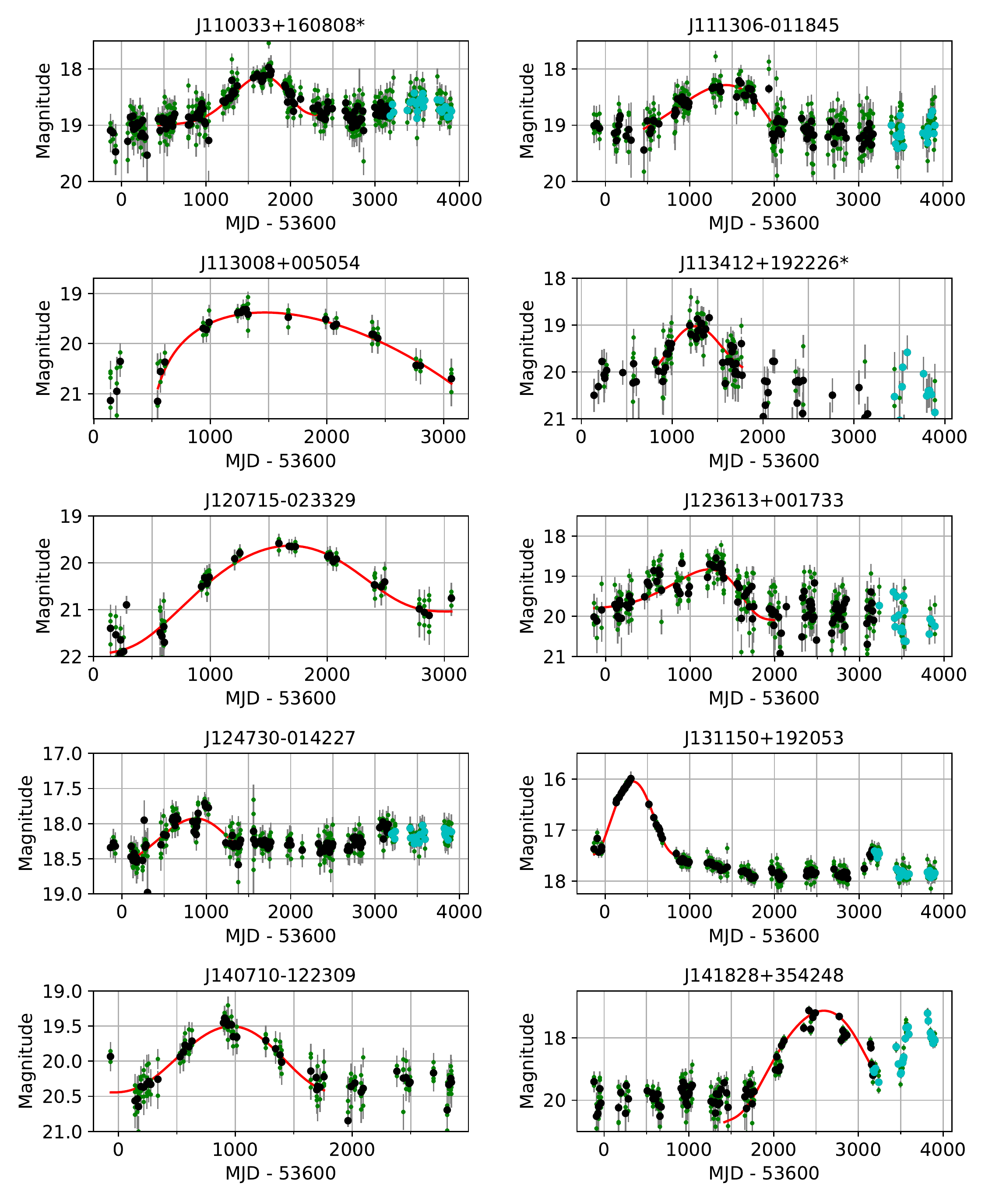}
\end{figure*}

\begin{figure*}
\centering
\contcaption{}
\includegraphics[width = 7.0in] {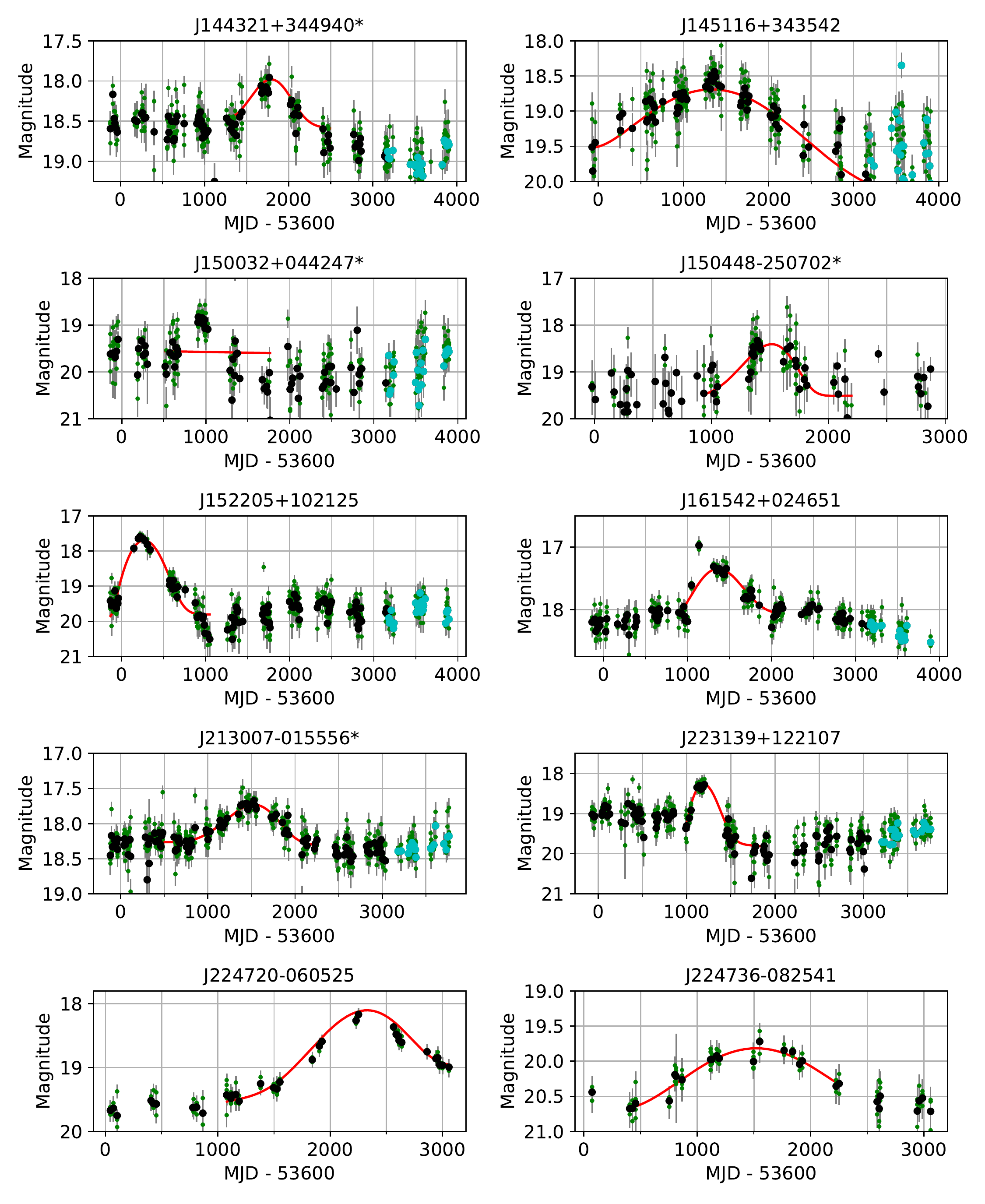}
\end{figure*}

\begin{figure*}
\centering
\contcaption{}
\includegraphics[width = 7.0in] {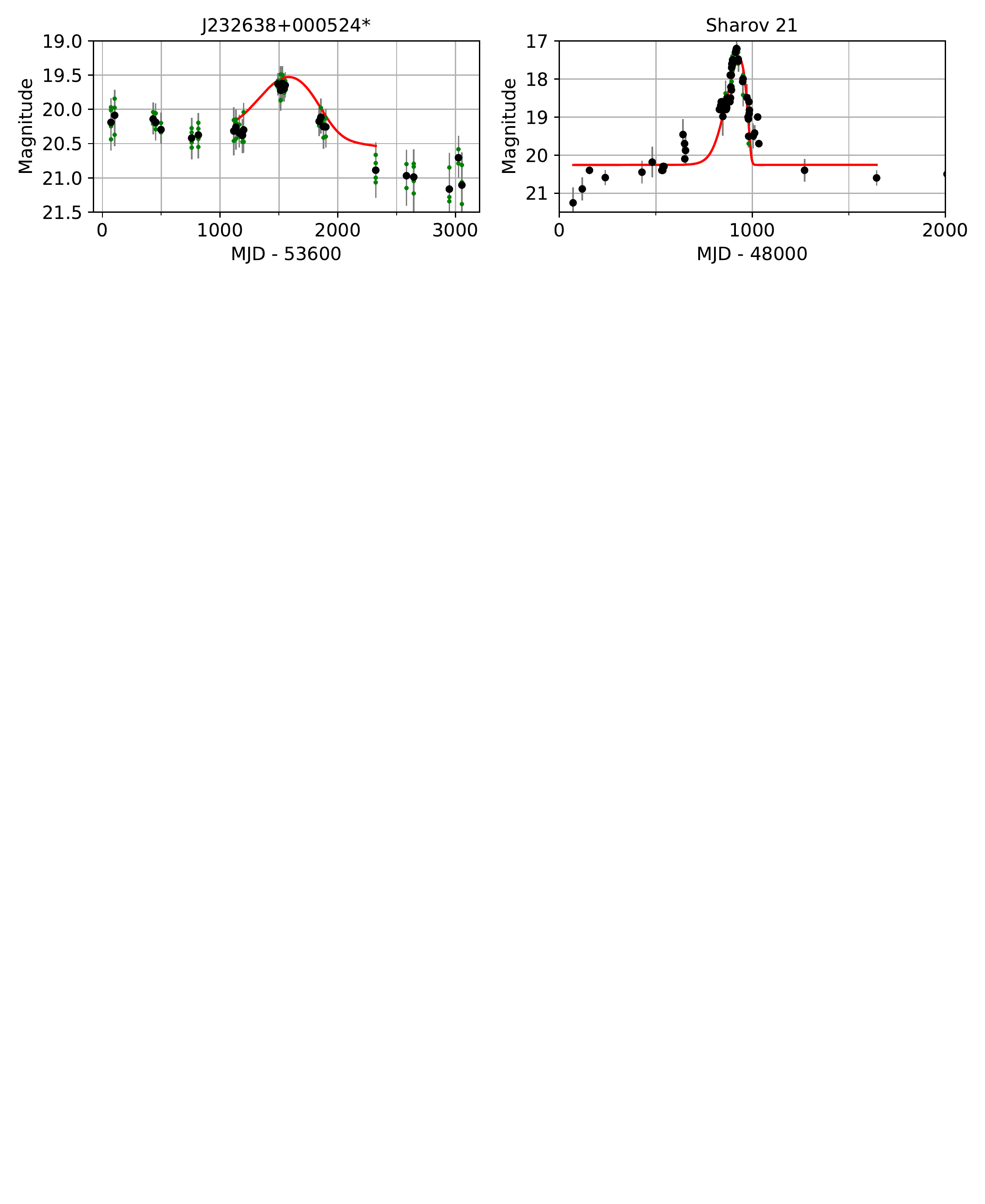}
\end{figure*}

\begin{figure*}
\centering
\caption{Best-fit single-point single-lens sources in order of increased reduced chi-square.}
\label{microlenslc}
\includegraphics[width = 7.0in] {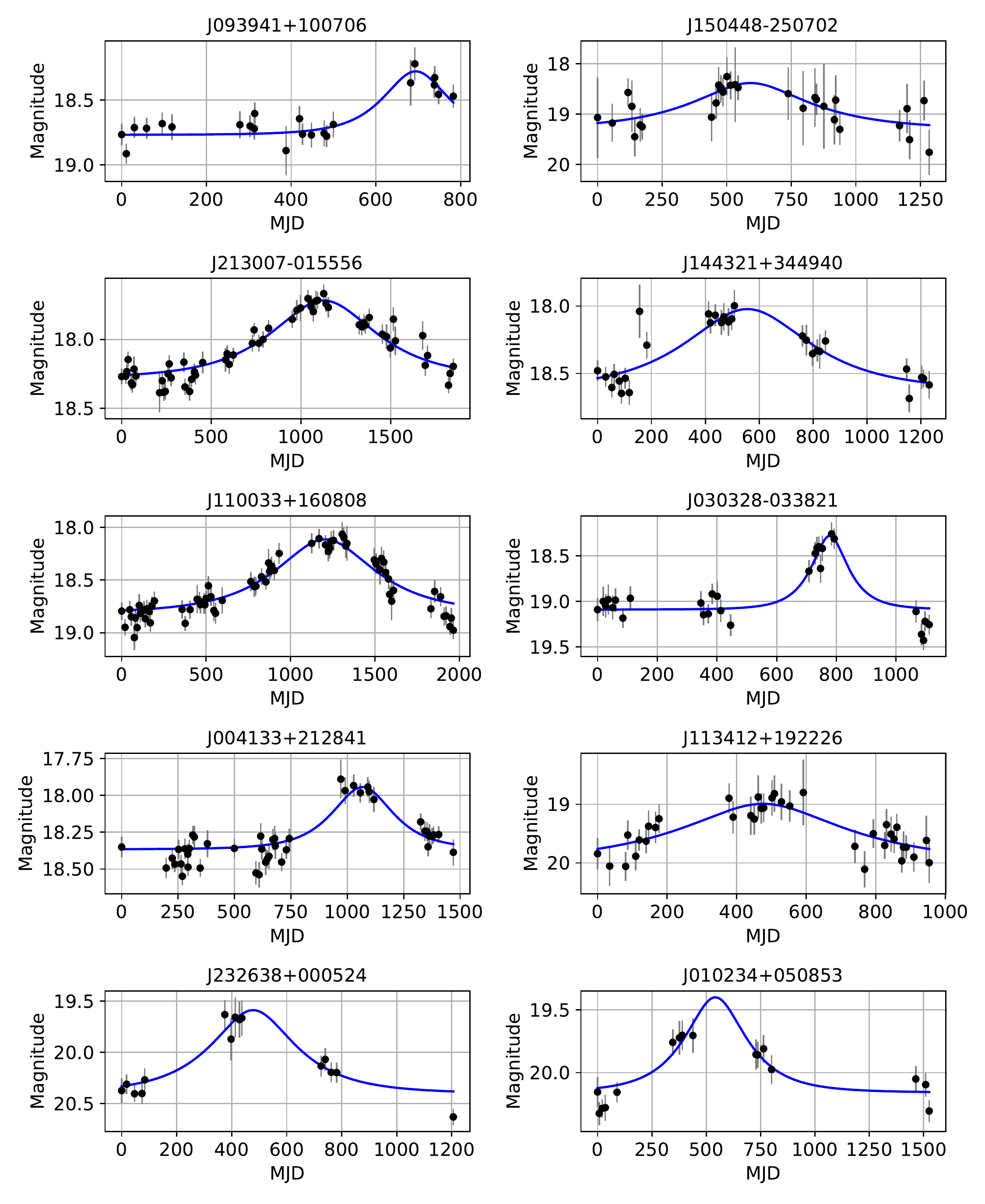}
\end{figure*}

\begin{figure*}
\centering
\contcaption{}
\includegraphics[width = 7.0in] {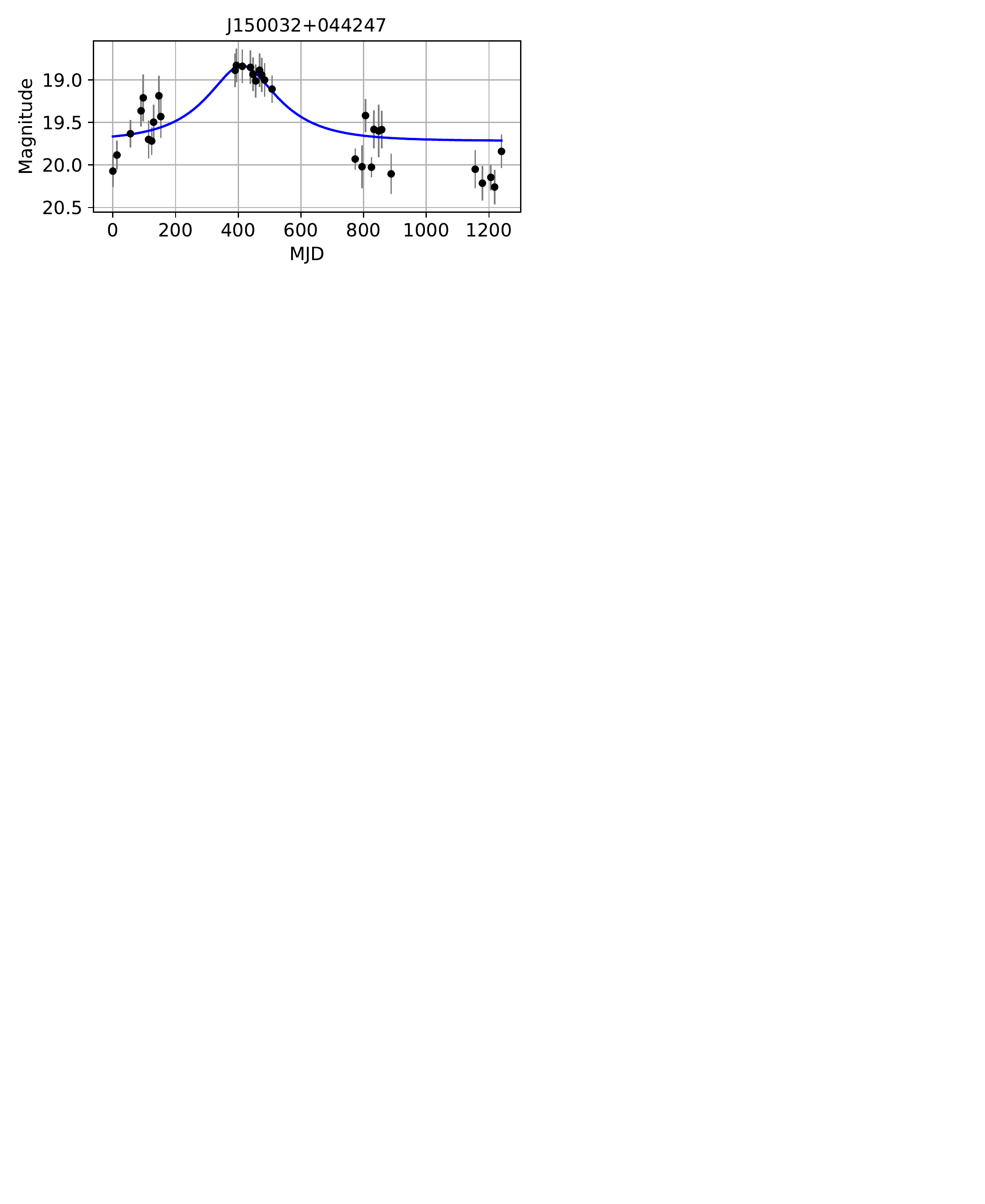}
\end{figure*}







\appendix
\section[]{Spectroscopic observations of flaring candidates}

Thirteen of the photometrically identified quasar candidates for which we found evidence of flaring activity did not have published spectroscopy (e.g., spectroscopic redshifts) prior to this work, hampering our ability to fully interpret the results for those sources.  Here we briefly describe spectroscopy obtained for these thirteen sources, which are indicated with asterisks after their redshifts in Table~\ref{candidates}.

\begin{figure*}
\centering
\caption{Examples of four optical spectra obtained from our Palomar and Keck follow-up programs.  The spectrum of CRTS~J224720.90-060525.8 was obtained at Palomar, while the other three spectra were obtained at Keck. Note the strong Fe features in the top panel flanking theH$\beta$ emission.}
\label{fig:spectra}
\includegraphics[angle=-90,width=7.0in]{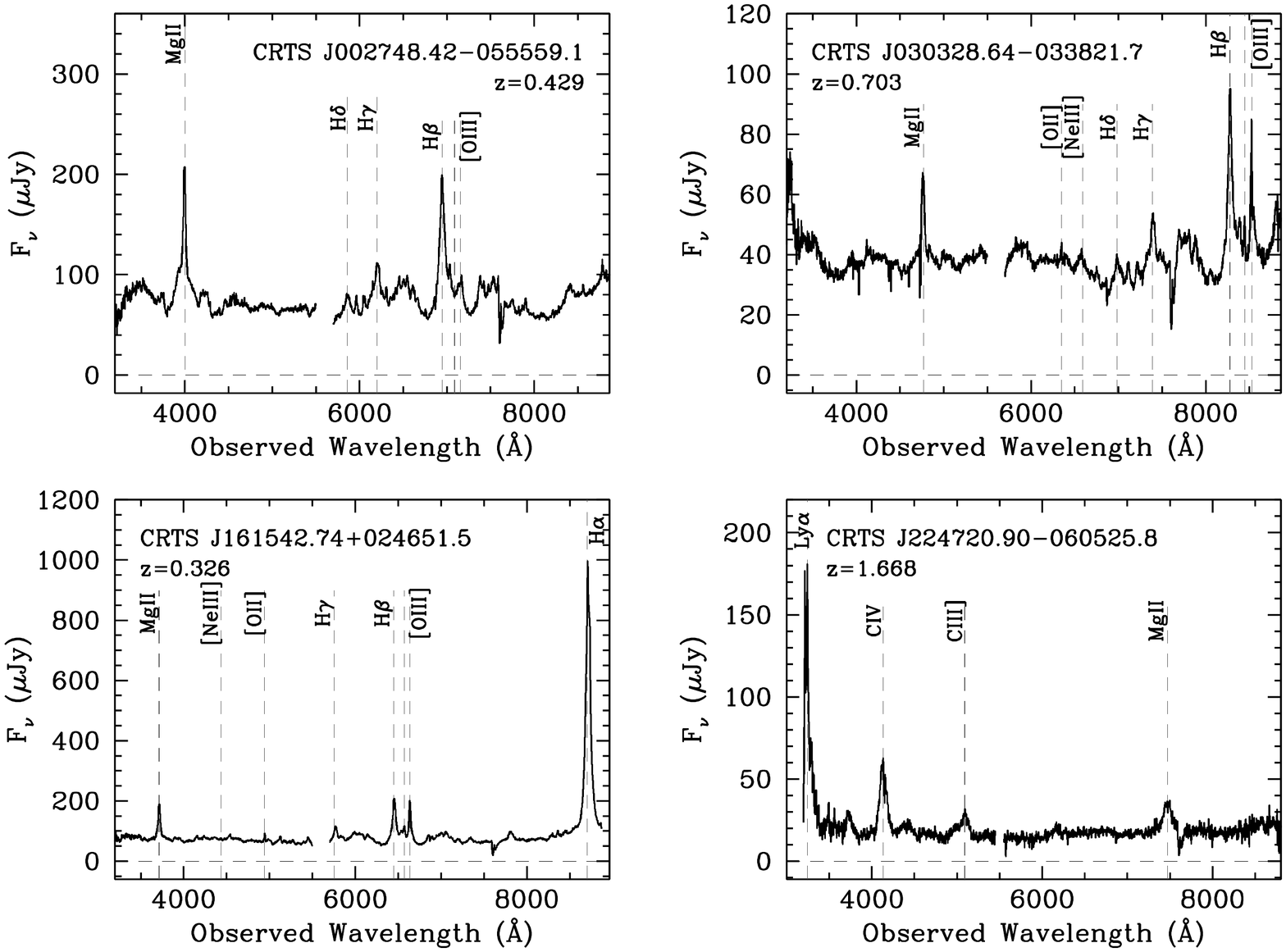}
\end{figure*}

We obtained spectroscopic observations at the Palomar and Keck Observatories between December 2015 and February 2017 from our dedicated program to follow-up CRTS AGN with unusual synoptic
properties (see Fig.~\ref{fig:spectra} for examples).  Table~\ref{spectra} lists basic observing details for the ten flaring quasars observed, including which telescope they were observed with, the date of the observation, and the integration time.  Palomar observations utilized the dual-beam Double Spectrograph on the 200-inch Hale Telescope, which was configured with the
1\farcs5 wide slit, the 5500~\AA\ dichroic, the 600 $\ell$ mm$^{-1}$ grating on the blue arm ($\lambda_{\rm blaze} = 4000$~\AA; spectral resolving power $R \equiv \lambda / \Delta \lambda \sim 1200$), and the 316 $\ell$ mm$^{-1}$ grating on the red arm ($\lambda_{\rm blaze} = 7500$~\AA; $R \sim 1800$).  Keck observations utilized the
dual-beam Low Resolution Imaging Spectrograph (LRIS: \cite{oke95}) on the Keck~I telescope, configured with the 1\farcs5 wide slit, the 600 $\ell$ mm$^{-1}$ grism on the blue arm ($\lambda_{\rm blaze} = 4000$~\AA; $R \sim 800$), and the 400 $\ell$ mm$^{-1}$ grating on the red arm ($\lambda_{\rm blaze} = 8500$~\AA; $R \sim 1000$).

All nights listed in Table~\ref{spectra} were photometric, and we processed the data using standard techniques within IRAF.  For all but the December 2015 observation, we calibrated the data using standard stars from \cite{stone83, stone84} and \cite{massey90} observed on the same nights using the same instrument
configuration; for the December 2015 observation, we used an archival sensitivity function obtained from similar standard stars observed with an identical instrument configuration. For all thirteen sources, the spectroscopy revealed quasars with multiple emission features providing robust redshift identifications.

\begin{table*}
\caption{Details of the 13 flaring quasars that were observed spectroscopically to derive redshifts.}
\label{spectra}
\centering
\begin{tabular}{llll}
\hline
ID               & Telescope & UT Date     & Exp. Time (s) \\
\hline
J000727$-$132644 & Keck      & 2016 Sep 09 & 900           \\
J002748$-$055559 & Keck      & 2016 Sep 09 & 900           \\
J005448$+$225123 & Palomar   & 2015 Dec 04 & 900           \\
J010234$+$050853 & Keck      & 2016 Sep 09 & 900           \\
J030328$-$033821 & Keck      & 2016 Sep 09 & 900           \\
J030606$+$192643 & Keck      & 2016 Sep 09 & 900           \\
J081333$+$183446 & Keck & 2016 Dec 29 & 900 \\
J090347$+$151818 & Palomar & 2017 Feb 25 & 900 \\
J090612$+$272347 & Palomar & 2017 Feb 25 & 900 \\
J161542$+$024651 & Keck      & 2016 Sep 09 & 900           \\
J223139$+$122107 & Keck      & 2016 Sep 09 & 900           \\
J224720$-$060525 & Palomar   & 2016 Nov 06 & 2$\times$900  \\
J224736$-$082541 & Palomar   & 2016 Nov 06 & 900           \\
\hline
\end{tabular}
\end{table*}

\section[]{Statistical moments of the translated Weibull distribution}
\label{weistat}

The translated Weibull distribution is defined as:

\[ X(p; a, s) = \left( \frac{a}{s} \right) p^{a-1}e^{-p^{a}} \]
\[ p = (t - t_0) / s\]

\noindent
in which $a$ is a shape parameter $(a > 0)$, the scale (or width) is specified by $s (s > 0)$, and the location is given by $t_0 (t_0 \ge 0)$; the independent coordinate is $t (t \ge t_0)$. The first three statistical moments, $\mu, \sigma^2, \gamma_1$, are:

\[ E(X^r) = \sum_{i = 0}^r  \binom{r}{i} t_0^is^{r-i} \Gamma\left(1 + \frac{r-i}{a}\right) \]

\[ \mu = E(X) = t_0 + s g_1 \]

\[ \sigma^2 = E(X^2) - \mu^2 = s^2 (g_2 - g_1^2) \]

\[ \gamma_1 = \frac{E(X^3) - 3 \mu \sigma^2 - \mu^3}{\sigma^3}  =  \frac{g_3 - 3 g_1 g_2 + 2 g_1^3}{(g_2 - g_1^2)^{3/2}} \]

\noindent
where $g_i = \Gamma(1 + i/a)$. The skewness of the Weibull distribution is invariant under a location-scale transformation with positive slope: 
$ skew(a + bX) = skew(X)$ where $b > 0$ and changes sign for a negative slope: $skew(a + bX) = -skew(X)$ for b < 0.

\bsp	
\label{lastpage}
\end{document}